\documentclass[Afour,sageh,times]{sagej}
\usepackage[ruled]{algorithm2e} 
\usepackage{moreverb,url}
\usepackage{algorithm2e}
\usepackage{rotating}
\usepackage{graphicx}
\usepackage{amsmath,amssymb,amsfonts}
\usepackage{booktabs}
\usepackage[colorlinks,bookmarksopen,bookmarksnumbered,citecolor=red,urlcolor=red]{hyperref}
\usepackage{xcolor}

\newcommand\BibTeX{{\rmfamily B\kern-.05em \textsc{i\kern-.025em b}\kern-.08em
T\kern-.1667em\lower.7ex\hbox{E}\kern-.125emX}}

\def\volumeyear{2024}

\begin{document}

\runninghead{GRAND-HC: Graph-Refined Author Name Disambiguation}

\title{GRAND-HC: Graph-Refined Author Name Disambiguation via Harmony Contrastive Learning}

\author{Yuanhao Sun\affilnum{1}\textsuperscript{*}, 
        Zhouyang Jin\affilnum{1}\textsuperscript{*}, 
        Yi Xu\affilnum{1}, 
        Luoyi Fu\affilnum{1}, 
        Jiaxin Ding\affilnum{1}, 
        Xiaoying Gan\affilnum{1}, 
        Xinbing Wang\affilnum{1}\textsuperscript{†}, 
        and Chenghu Zhou\affilnum{2}}

\affiliation{\affilnum{1}Shanghai Jiao Tong University, Shanghai, China\\
             \affilnum{2}Institute of Geographic Sciences and Natural Resources Research, Chinese Academy of Sciences, Beijing, China}

\corrauth{Xinbing Wang, Shanghai Jiao Tong University, 800 Dongchuan Road, Shanghai, 200240, China.}

\email{xwang8@sjtu.edu.cn}

\begin{abstract}
From-Scratch Name Disambiguation (SND), a core task of Author Name Disambiguation (AND), aims to group papers sharing an identical ambiguous name into clusters corresponding to distinct real-world authors. {However, existing SND methods suffer from two critical, largely overlooked limitations: first, the inherent long-tailed uneven author class distribution — where most papers under an ambiguous name belong to a small number of prolific authors — severely biases representation learning, leading to low-discriminative embeddings and over-merging of less-published tail authors; second, existing cluster number estimation methods are unreliable and poorly scalable for long paper sequences, restricting end-to-end SND deployment in real-world large-scale scenarios. To address these issues, we propose GRAND-HC, a complete end-to-end SND framework with three targeted components. We first construct a heterogeneous paper graph based on co-author, co-organization and co-venue relations, and adopt a graph attention network as the embedding generation backbone. Then, harmony contrastive learning (HCL) dynamically reweights training loss to suppress overfitting to prolific authors, learning highly discriminative embeddings with clear boundaries between tail authors. On this basis, a graph-refined distance matrix (GRDM) leverages graph topology to adaptively optimize pairwise distances, further preventing over-merging of tail authors in clustering. Meanwhile, a lightweight Paper Compression Module (PCM) achieves accurate and robust cluster number estimation across varying paper scales, eliminating the long-sequence modeling defect of existing methods.} Finally, Hierarchical Agglomerative Clustering outputs the final author clusters with the optimized distance matrix and estimated cluster number. Extensive experiments on multiple benchmarks demonstrate that GRAND-HC outperforms state-of-the-art models on macro F1 score. Furthermore, GRAND-HC has been successfully deployed in a billion-scale academic database, with source code released in https://github.com/baokou-fw2/GRAND-HC.
\end{abstract}

\keywords{Author Name Disambiguation, graph-refined distance matrix, Harmony Contrastive Learning, Cluster Size Estimation}

\maketitle
\footnotetext{\textsuperscript{*}Both authors contributed equally to this research.}
\footnotetext{\textsuperscript{†}Corresponding author.}

\section{Introduction}
The author name disambiguation (AND) problem i.e. authors sharing the same name, is a critical challenge in scientific literature database management and information retrieval such as AMiner~\cite{tang2008arnetminer} and DBLP~\cite{boukhers2024deep}. This problem has long been commonly divided into three sub-tasks~\cite{chen2023web}: From-Scratch Name Disambiguation (SND), Real-time Name Disambiguation, and Incorrect Assignment Detection. In this paper, we focus on From-Scratch Name Disambiguation (SND) problem, which aims to group all papers by the same author for building a new academic system from the ground up.

\begin{figure}[htb]
  \centering
  \includegraphics[width=\linewidth,height=6cm]{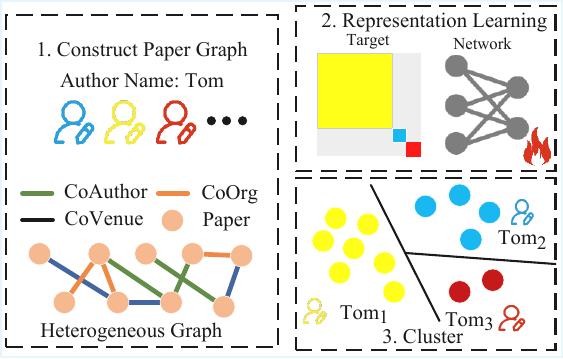}
  \caption{General flow of current SND frameworks.}
  \label{flow}
\end{figure}

Addressing the From-Scratch Name Disambiguation (SND) task requires effectively aggregating multi-source heterogeneous information from academic datasets to distinguish real authors with identical names, which is a fundamental component for academic knowledge graph construction, scholar evaluation, and academic search. As illustrated in Figure~\ref{flow}, existing mainstream SND frameworks follow a three-stage general paradigm for a given ambiguous name (e.g., \textit{`Tom'} with a pile of associated papers). First, they construct a heterogeneous paper graph, where nodes represent individual papers, and edges encode diverse relational clues including co-author, co-venue, and co-institution affiliations~\cite{cappelli2025recent,zhang2018name}. Second, to learn discriminative paper embeddings, supervised methods construct a target paper relation matrix from labeled data to guide network training~\cite{liu2024author,chen2020conna, zhou2024towards}, while unsupervised counterparts leverage intrinsic paper correlations as pseudo supervision signals~\cite{cheng2024bond, santini2022knowledge}. Finally, based on the learned embeddings, they perform clustering with an estimated number of author clusters, and assign papers to distinct real authors (e.g., Tom$_1$, Tom$_2$, and Tom$_3$ in Figure~\ref{flow}).

{Despite the promising performance of existing frameworks, they still face two critical limitations that severely hinder their performance and generalization in real-world scenarios. }

{\textbf{Limitation 1 in the representation learning stage (Step 2 in Figure~\ref{flow}): most existing SND methods largely overlook the inherent long-tailed uneven distribution of papers across real authors, which severely undermines the core goal of this stage — learning highly discriminative paper embeddings, and remains a ubiquitous unaddressed challenge in real-world academic datasets.}}

\begin{figure}[htb]
  \centering
  \includegraphics[width=\linewidth]{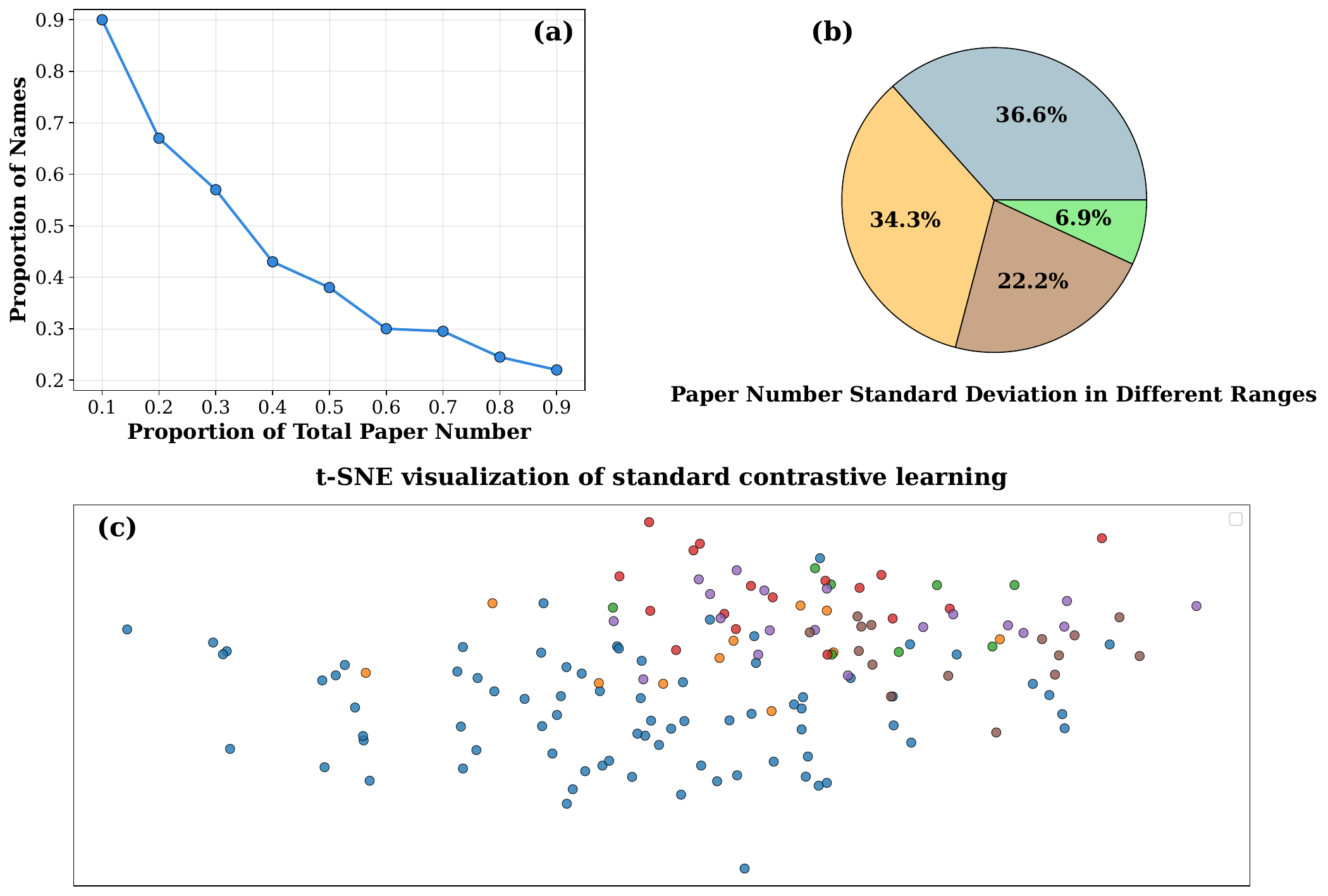}
  \caption{Statistical analysis of the long-tailed data distribution in SND tasks and its negative impact on representation learning. (a) Proportion of ambiguous names containing at least one prolific author whose papers account for x\% of the total papers under the same name. (b) Distribution of the standard deviation of paper counts across distinct authors within the same ambiguous name. {(c) t-SNE visualization of paper embeddings learned by standard contrastive learning.}}
  \label{motivation1}
\end{figure}

For a given ambiguous name, a small number of prolific high-volume authors hold the majority of associated papers. To quantitatively validate the ubiquity of this phenomenon, we conduct a statistical analysis on two widely used large-scale SND benchmarks: {WhoisWho-SND\footnote{\url{https://github.com/THUDM/WhoIsWho}} and AMiner-v2\footnote{\url{https://github.com/neozhangthe1/disambiguation}}}, both constructed from real-world academic literature. As shown in Figure~\ref{motivation1}(a), nearly 40\% of ambiguous names contain a dominant prolific author that accounts for more than 50\% of the total papers under the name. Figure~\ref{motivation1}(b) further quantifies the dispersion of paper counts: for 29.1\% of ambiguous names, the standard deviation of paper counts across different authors exceeds 50, reflecting an extremely uneven distribution with high volatility relative to the mean. These statistics confirm that the uneven paper distribution is an inherent and widespread characteristic of real-world SND tasks. {This severe data imbalance brings fatal defects to existing frameworks. Conventional representation learning strategies are inherently biased towards dominant prolific authors during training, resulting in a low-discriminative embedding space where tail authors' papers are heavily overlapped with no clear clustering boundaries, as visualized in Figure~\ref{motivation1}(c).}

{\textbf{Limitation 2 in the final clustering stage (Step 3 in Figure~\ref{flow}): existing methods suffer from unreliable cluster number estimation, which is an indispensable prerequisite for accurate end-to-end author clustering (the core task of this stage), and severely restricts the practical deployment of SND frameworks in large-scale real scenarios.}}

{Mainstream cluster number estimation approaches for SND can be divided into two categories, both with inherent fatal flaws. For distance-based methods~\cite{cheng2024bond,huang2025framework}, they rely on hand-crafted distance thresholds to determine cluster numbers, which are extremely sensitive to data distribution shifts and embedding quality fluctuations, leading to severe systematic estimation bias.}

{For RNN-based methods widely adopted in prior SND works~\cite{zhang2018name}, they fail to model overlong paper sequences effectively due to the inherent gradient vanishing problem of recurrent architectures~\cite{hochreiter1997long}, thus cannot capture the global structural information of the heterogeneous paper graph when the number of papers grows large.}

\begin{figure}[htb]
  \centering
  \includegraphics[width=\linewidth]{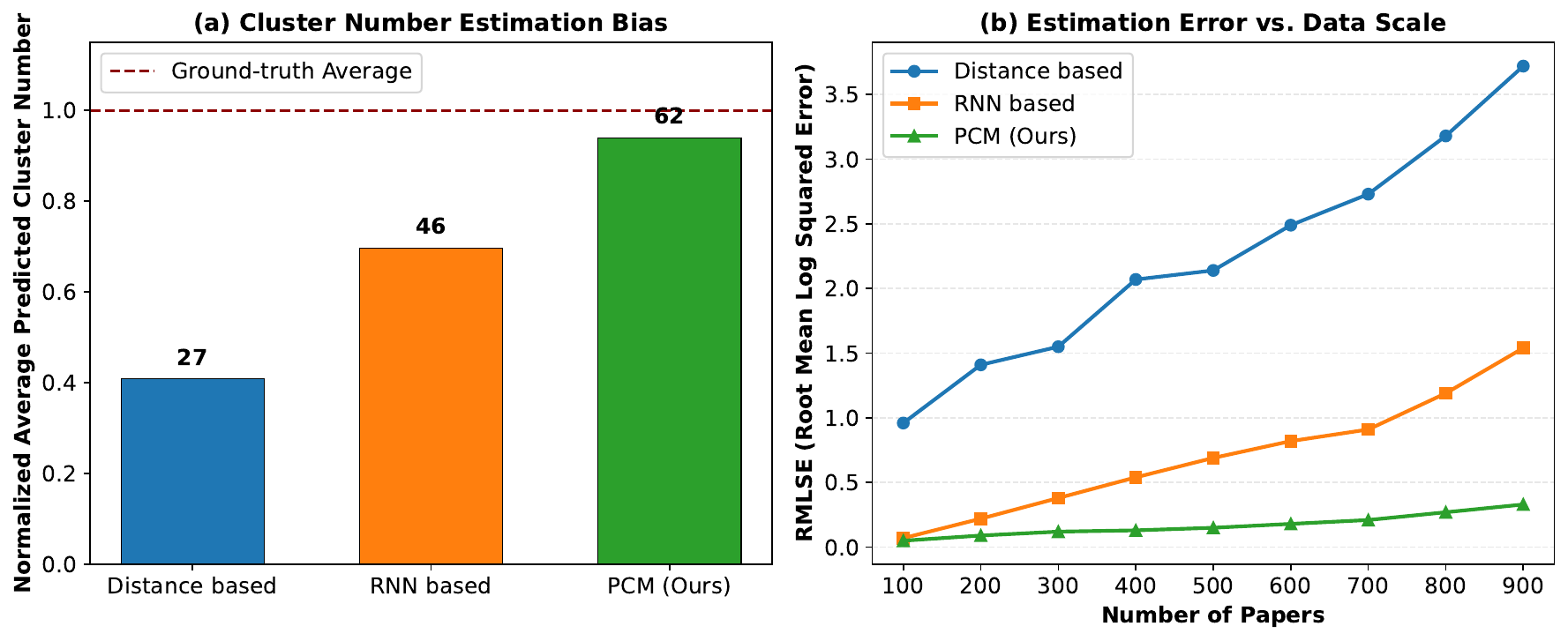}
  \caption{{Quantitative evaluation of cluster number estimation performance on the AMiner-v2 dataset. (a) Normalized average predicted cluster number of different methods, where the red dashed line denotes the ground-truth average cluster number. (b) Estimation error (RMLSE) of different methods with the increasing number of papers.}}
  \label{motivation2}
\end{figure}

{To quantitatively verify the ubiquity of these defects, we conduct comparative experiments on the widely used AMiner-v2 dataset, with results shown in Figure~\ref{motivation2}. As illustrated in Figure~\ref{motivation2}(a), mainstream methods suffer from severe underestimation of the real cluster number: the average predicted cluster number of distance-based methods is only 27, and that of RNN-based methods is 46, both far below the ground-truth average of 62. Figure~\ref{motivation2}(b) further demonstrates the poor scalability of existing methods: as the number of papers increases from 100 to 900, the Root Mean Log Squared Error (RMLSE) of distance-based methods surges from 0.96 to 3.72, while the estimation error of RNN-based methods rises sharply from 0.07 to 1.24. Without reliable and robust cluster number guidance, even high-quality paper embeddings cannot produce accurate end-to-end disambiguation results, which severely limits the deployment of existing SND frameworks in real-world large-scale academic scenarios.}

{To address the two aforementioned critical limitations in the standard three-stage SND pipeline (Figure~\ref{flow}), we propose GRAND-HC, a unified end-to-end from-scratch name disambiguation framework, where each component is explicitly designed to target one well-identified limitation.}

{To address \textbf{Limitation 1}, we design two tightly coupled, progressive components to eliminate prolific bias and over-merging across the full pipeline: \textit{Harmony Contrastive Learning (HCL)} and \textit{Graph-Refined Distance Matrix (GRDM)}. During each training epoch, we first run HAC on the current embeddings to dynamically stratify samples: correctly clustered pairs are mostly easy samples from prolific authors, while misclustered pairs are hard samples from under-represented tail authors. HCL then adaptively up-weights the loss of these hard samples, reversing the training signal bias, suppressing overfitting to prolific authors, and learning highly discriminative embeddings with clear boundaries for tail authors.}
{On the basis of the optimized embeddings, we further design GRDM to eliminate residual over-merging risks in the clustering stage. GRDM first refines the biased initial graph topology via embedding similarity, then adaptively adjusts the pairwise distance matrix according to node degree and edge connectivity. This fully leverages heterogeneous graph structural information to strengthen clustering boundaries, fixing the core defect of conventional distance-based clustering under long-tailed data distribution.}

{For \textbf{Limitation 2} in the final clustering stage (Step 3), we propose a novel Paper Compression Module (PCM) for accurate and robust cluster number prediction. Specifically, PCM first uses fixed-length learnable parameters as queries, and the learned paper embeddings as keys and values, to build a cross-attention module. This compresses the massive long paper sequence into a compact fixed-length sequence, retaining the most critical information related to cluster number estimation while eliminating the long-sequence modeling difficulty of traditional RNN-based methods. The subsequent Bi-LSTM then processes the compressed high-quality sequence, achieving significantly better generalization performance and estimation accuracy across different paper scales, even for large-scale long-sequence inputs where existing methods fail in Figure~\ref{motivation2}.}

The main contributions of this paper are summarized as follows:
\begin{itemize}
\item { We identify and quantitatively verify two core, largely overlooked limitations in the standard SND pipeline: (1) the long-tailed uneven paper distribution skews representation learning and causes severe over-merging of tail authors; (2) existing cluster number estimation methods suffer from poor scalability and unreliable performance as the number of papers grows.}
\item {To address the first limitation, we design two tightly coupled components: harmony contrastive learning (HCL) to alleviate model bias towards prolific authors and learn discriminative embeddings, and a graph-refined distance matrix (GRDM) to strengthen clustering boundaries and avoid over-merging of tail authors.}
\item {To address the second limitation, we propose a novel Paper Compression Module (PCM) for accurate and robust cluster number estimation, which eliminates the long-sequence modeling defect of traditional methods and maintains stable performance across different data scales.}
\item Extensive experiments show that GRAND-HC achieves state-of-the-art (SOTA) performance on multiple classic SND benchmarks. Furthermore, the framework has been successfully deployed in a billion-scale academic database for real-world SND tasks, verifying its effectiveness and practicality.
\end{itemize}

{The rest of the paper is organized as follows. 
In the related work section, we provide a literature review for SND. 
We formally formulate the SND problem in the problem formulation section. 
In the methodology section, we present the details of GRAND-HC. 
In the experiments section, we present our experimental results on several classic SND datasets and a comprehensive ablation study. 
Finally, we make conclusions in the conclusion section.}

\section{Related Work}
\label{relatedwork}
This section primarily delves into research concerning SND task, which is typically regarded as a clustering problem. The SND task involves two closely related components: paper representation learning and clustering. Current methods typically focus on learning paper representations to compute similarity, followed by clustering to assign papers to correct authors. Paper representation learning methods can be broadly categorized into feature-based and graph-based methods. In terms of clustering techniques, previous methods group data via distance matrix computed from paper embeddings, facing the challenge of unknown author cluster size.

\subsection{Feature-based Methods}
There are some methods generate embeddings mainly based on paper information with less consideration about graph structure~\cite{silva2017feature, han2004two, huang2006efficient,wang2020author,santini2022knowledge,zhang2024enhancing,xiao2021oag}. {\cite{han2017semantic} proposes a semantic fingerprint-based approach for author name disambiguation in Chinese documents, integrating text fingerprints with co-author and institution features to address name ambiguities. \cite{louppe2016ethnicity} presents a semi-supervised learning-based model, incorporating phonetic blocking strategies and ethnicity-sensitive features to build a linkage function, combined with hierarchical agglomerative clustering for efficient document grouping. \cite{wang2020author} proposes a framework based on adversarial representation learning, integrating content and relational information through heterogeneous information networks, while employing a self-training strategy and a random walk-based generation algorithm to handle high-order connections. \cite{santini2022knowledge} proposes a multimodal knowledge graph embedding framework based on LiteralE, integrating literal features (e.g., titles, dates) into entities. Two variants are introduced: LAND-glin, employing linear projection of title embeddings, and LAND-ggru, which fuses textual and numeric literals via GRU. Recently, with the advancement of large language models, \cite{zhang2024enhancing} fine-tunes several large language models to collaboratively address the IAD task. However, in the context of SND, due to the vast volume of literature, there appears to be limited progress so far in developing LLM-based approaches. In contrast, \cite{xiao2021oag} trains the BERT model on academic text corpus in Open Academic Graph, including paper titles, abstracts and bodies and finetune it in SND tasks, which achieves outperformance over many baselines.}

\subsection{Graph-based Methods}
With the advancement of graph representation learning, Graph Neural Networks (GNNs) and their variants are increasingly used for extracting relational information from heterogeneous paper-author graphs~\cite{shin2014author, sun2020pairwise, fan2011graph, santini2022knowledge, bekkerman2005disambiguating, hermansson2013entity, kanani2007improving,cheng2024bond,liu2024author}. {\cite{chen2021name} proposes a Graph Convolutional Network (GCN) model that integrates attribute features and linkage information from various graphs, using hierarchical clustering for disambiguation. \cite{ma2020graph} combines a Variational Graph Auto-Encoder (GAE) with graph embedding to capture attribute and topological features, employing hierarchical agglomerative clustering for document partitioning. \cite{pooja2022exploiting} leverages a self-attention-based graph convolution network on multi-hop neighborhoods, incorporating neighborhood and relation-level attention to improve document embeddings and clustering accuracy. \cite{xiong2021learning} proposes a joint representation learning framework that simultaneously embeds semantic and relationship information into a low-dimensional space using a Variational AutoEncoder (VAE) and applies HAC for SND problem. \cite{cheng2024bond} proposes an end-to-end framework, BOND, which constructs multi-relational graphs, leverages Graph Attention Networks (GAT) for local metric learning, and utilizes DBSCAN clustering to jointly optimize local and global signals. \cite{liu2024author} proposes a framework combining paper association graph refinement and contrastive learning, dynamically optimizing graph structures to reduce noise and uncertainties, while employing multi-level contrastive learning to enhance the discriminative power of semantic and structural representations. \cite{gong2024more} integrates OAG-BERT, SimCSE, LightGBM, and iHGAT via supervised and unsupervised learning.}

\subsection{Clustering Size Estimation for SND Task}
In practical SND tasks, the true number of authors is often unknown, making the estimation of the number of clusters a crucial factor for the final disambiguation results. Although a segment of the methods focuses on directly inferring text-pair relationships, it lacks the ability to generate distinct paper clusters and suffers from notable inefficiency in computation~\cite{zhou2021multiple,ji2018machine,zhou2024towards}. The main current algorithms for category number estimation fall into four categories: Bayesian-based estimation~\cite{ pelleg2000extending}, clustering algorithms using predefined thresholds~\cite{ cheng2024bond}, graph-based clustering algorithms~\cite{tang2011unified,han2005name,qiao2019unsupervised} and neural network-based algorithms~\cite{zhang2018name}. {\cite{ pelleg2000extending} iteratively splitting the centroids and searching for an optimal K based on the quality of the proposes clustering. \cite{cheng2024bond} Classless number clustering using DBSCAN clustering algorithm with predefined thresholds. \cite{tang2011unified} proposes a unified probabilistic framework, employing a dynamic method to estimate the number of clusters \textit{K}. \cite{han2005name} employs a K-way spectral clustering method for name disambiguation in author citations, predefining the number of clusters \textit{K} as the labeled ground truth, focusing on partitioning citations into clusters that correspond to unique authors. \cite{qiao2019unsupervised} utilizes an optimal modularity partitioning mechanism to determine the partition of publications, which iteratively calculates modularity after each cluster merging process until there is no edge among clusters, and choose the largest M as the final clustering result. \cite{zhang2018name} proposes an RNN-based neural network model and construct a training set with strictly defined ranges for the total number of papers and the number of author categories to mitigate the limitations of RNNs in processing long sequences.}

\section{Problem Formulation}
\label{problem-definition}
In this section, we introduce the foundational concepts of author name disambiguation and formally define the SND problem.

\textit{Definition 3.1.} \textbf{Paper}. A paper $p$ possesses multiple attributes, denoted as $p = \{x_1, \cdots, x_F\}$, where $x_f \in p$ denotes the $f$-th attribute (e.g., title) and $F$ is the number of attributes. These attributes include, but are not limited to, the title, abstract, keywords, authors and their affiliated organizations, the publishing journal or conference, and the publication date.

\textit{Definition 3.2.} \textbf{Author}. An author $a$ has a collection of papers they have written, denoted as $a = \{p_1, \cdots, p_n\}$, where $n$ is the number of papers authored by $a$.

\textit{Definition 3.3.} \textbf{Candidate Papers}. Given an author name $na$, the candidate papers consists of all papers where one of the authors is named $na$, defined as $\mathcal{P}^{na} = \{p_1^{na}, \dots, p_N^{na}\}$.

\textit{Problem} \textbf{From-scratch Name Disambiguation (SND)}. Given an author name $na$ and its corresponding candidate papers $\mathcal{P}^{na}$, the goal of SND is to find a function $\Phi$ that partitions $\mathcal{P}^{na}$ into multiple disjoint subsets of papers $\mathcal{C}^{na} = \{\mathcal{C}_1^{na}, \mathcal{C}_2^{na}, \dots, \mathcal{C}_K^{na}\}$. Each subset $\mathcal{C}_i^{na}$ should exclusively contain papers authored by the same author named $na$, while papers in different subsets should belong to distinct authors. The parameter $\mathcal{K}$ represents the total number of distinct authors sharing the name $na$. This process can be viewed as a clustering problem, formally defined as follows:
\begin{equation}
    \Phi(\mathcal{P}^{na}) \rightarrow \mathcal{C}^{na}, \text{ where } \mathcal{C}^{na} = \{\mathcal{C}_1^{na}, \mathcal{C}_2^{na}, \dots, \mathcal{C}_\mathcal{K}^{na}\}.
\end{equation}
The key of the problem is to find the accurate $\mathcal{P}^{na}$ representation and clustering number $\mathcal{K}$, and then divide the paper using a suitable clustering algorithm $\Phi$.

\section{Methodology}
\label{GEDCL}
\begin{figure*}[htbp]
  \centering
  \includegraphics[width=\textwidth]{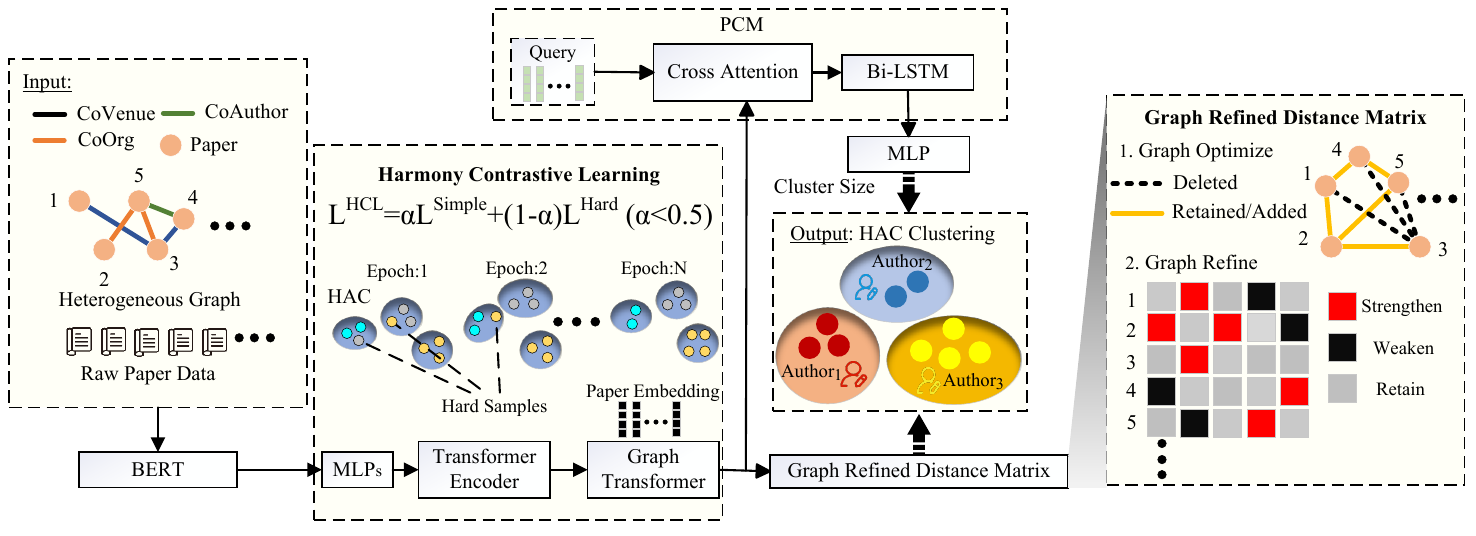}
  \caption{The overall framework of GRAND-HC. We first construct a heterogeneous graph based on paper relations and get the initial semantic embedding by BERT encoding. Then, representation network is trained by harmony contrastive learning and outputs paper embedding. With these paper embeddings, we train cluster size estimation network and calculate graph-refined distance matrix. Finally, HAC clusters papers into different authors with specified cluster size and distance matrix.}
  \label{framework}
\end{figure*}

In this section, we provide a detailed explanation of GRAND-HC. We first describe our method for constructing heterogeneous graph. Then, we introduce paper embedding generation network, {which is trained by harmony contrastive learning(HCL)}. Then, we calculate graph-refined distance matrix(GRDM) and estimate cluster size with paper compression module(PCM) for HAC clustering. Finally, the analysis of complexity and algorithm workflow is summarized. The overall framework is illustrated in Figure~\ref{framework}.

\subsection{Paper Relational Heterogeneous Graph Construction}
\label{paper relation}
To better leverage the relationships between papers, we constructed a heterogeneous paper relationship network for candidate papers under each author name. Each connected component of the network represents a potential article set of a single author. In this network, nodes represent papers, whose attributes include features required for subsequent semantic information extraction (e.g., title, abstract, and keywords). Edges represent the similarity or relationships between papers and are the key to constructing the heterogeneous network.

\subsubsection{Edge Construction}
The relational information of papers serves as a crucial feature for measuring similarity and is used to establish edges between papers. Traditional studies~\citep{zhang2017name, tang2011unified} often rely on co-authorship and citation relationships to measure paper similarity. This study categorizes relationships into the following three types:

\begin{itemize}
    \item \textbf{Co-author Relationship}: Overlap of authors other than the target disambiguated author $na$.
    \item \textbf{Co-organization Relationship}: Overlap in one or more organizations of the target disambiguated author $na$.
    \item \textbf{Co-venue Relationship}: Papers published in the same conference or journal.
\end{itemize}

The detailed construction of these different types of edges will be elaborated in section experiments.

\subsubsection{Text Similarity Measurement}
To measure the similarity of different attributes between papers, we employed several text-matching methods:

\begin{itemize}
    \item \textbf{Co-author}: The identification process relies on name matching. To address inconsistencies, preprocessing eliminates the effects of hyphens and spaces. Segmented matching handles variations in name order, while the edit distance algorithm corrects abbreviations and minor spelling errors.
    \item \textbf{Co-organization}: The name matching algorithm first determines the affiliation of the disambiguated author (\textit{na}). To refine granularity inconsistencies across university, department, and geographic levels, the Jaccard index is applied for more precise matching.
    \item \textbf{Co-venue}: To ensure accurate co-venue identification, redundant information in publication venue data---such as publication years and name abbreviations---is filtered out. The Jaccard index is then used to quantify similarity.
\end{itemize}

\subsection{Paper Embedding Generation}

Firstly, BERT takes paper title, abstract, venue, authors, organizations and keywords and outputs 768 dimensions paper embedding. In addition, previous methods neglect certain SND-related graph information, such as the high likelihood that papers within the same connected component are authored by the same individual. As a result, we code every connected component and give each paper a subgraph label according to its location.

Suppose we have a paper relational heterogeneous Graph $G = (V, E)$, where $V$ is the set of papers and $E$ is the set of edges (denoting the relationships). We first find all the connected components of the graph $G$. Let $C = \{C_1, C_2, \ldots, C_k\}$ be the set of all connected components of the graph $G$, where $C_i$ denotes the $\text{i-th}$ connected component and $k$ is the total number of connected components. For each paper $v \in V$, we define its subgraph label as:
{\begin{equation}
\label{subgraphlabel}
    \text{subgraph label}_{v}=i, \quad i \in \{1, 2, \ldots, k\}
\end{equation}}
where $i$ denotes that the paper $v$ belongs to the connected component $C_i$.
To combine the graph information with semantic representation from BERT, we first use four independent MLPs to transform them into the same dimension:
\begin{equation}
\label{MLP}
  \mathbf{Y^{k}}=MLP^{k}(\mathbf{X^{k}}),
\end{equation}
where k from 0-3 corresponds to paper initial semantic embedding from BERT, subgraph label, degree and year. Then, we use $\mathbf{Y^{0}}$ to aggregate them with a transformer encoder:
\begin{equation}
\label{transformer}
\mathbf{P} = \text{Transformer Encoder}([\mathbf{Y^{0}},\mathbf{Y^{1}},\mathbf{Y^{2}},\mathbf{Y^{3}}])[0].
\end{equation}

{Then, we utilize a variant of the Graph Attention Network (GAT)~\cite{velivckovic2017graph} proposed by~\cite{shi2020masked} as a structure encoder to get final paper embedding $\mathbf{H}$ and calculate similarity matrix $\mathbf{S}$:}
\begin{equation}
\label{graphtransformer}
  \textbf{H} = Graph Attention(\textbf{P},\textbf{A}), \quad \textbf{S} = \left[\frac{<H_{i},H_{j}>}{\parallel H_{i}\parallel \parallel H_{j}\parallel}\right]^{N\times N},
\end{equation}
where $\mathbf{A}$ is the original adjacency matrix of $\mathbf{G}$ and $\textbf{S}$ is the cosine similarity matrix $(S_{ij}\in [-1,1])$.

\subsection{Harmony Contrastive Learning (HCL)}
{The core goal of the representation learning stage in SND is to learn a highly discriminative embedding space where papers from the same author are close while papers from different authors are far apart. However, as identified in Limitation 1, the inherent long-tailed uneven paper distribution severely undermines this goal. }

Standard contrastive learning methods for SND~\cite{cheng2024bond,zhou2021multiple} fail to address this issue, as they minimize the standard binary cross-entropy (BCE) loss between the ground-truth paper relationship matrix $\mathbf{L}$ and the cosine similarity matrix $\mathbf{S}$ of learned embeddings:
\begin{align}
\label{sim}
  \textbf{L} &= [\text{L}_{i}=\text{L}_{j}]^{N\times N},
\end{align}
\begin{equation}
\label{BCE}
\begin{split}
BCE(\textbf{S},\textbf{L}) = -\frac{1}{N}\sum_{i=1}^N \sum_{j=1}^N 
[\sigma(S_{ij}) \log(\sigma(L_{ij})) + \\
(1 - \sigma(S_{ij})) \log(1 - \sigma(L_{ij}))],
\end{split}
\end{equation}
where $\text{L}_{i}$ denotes the unique author that paper $i$ belongs to in the ground truth, and $\sigma$ is the Sigmoid function. {The critical defect of this formulation is that the training signal is completely dominated by paper pairs from prolific authors: since prolific authors hold the majority of papers, most positive/negative pairs in Eq.~(\ref{BCE}) are correlated with them. This causes the model to overfit the research topics and collaboration patterns of prolific authors, resulting in an embedding space where tail authors' papers are severely overlapped with extremely low discriminability, which directly leads to the over-merging issue in the subsequent clustering stage.}

{To address this problem and learn a balanced, high-discriminative embedding space (the core requirement for solving Limitation 1), we propose Harmony Contrastive Learning (HCL), which dynamically reweights the training loss to suppress the overfitting to prolific authors and force the model to focus on learning discriminative features for tail authors. The key insight is that paper pairs from prolific authors are usually "simple samples" that are easy to cluster correctly, while pairs from tail authors are "hard samples" that contain critical discriminative information but are masked by the dominant simple samples.}

Specifically, during each training epoch, we first perform Hierarchical Agglomerative Clustering (HAC) on the current learned embeddings with the ground-truth cluster number $\mathcal{K}$, and construct a predicted paper relationship matrix based on the HAC results:
\begin{equation}
    \textbf{Pre} = [\text{Pre}_{i}=\text{Pre}_{j}]^{N\times N},
\end{equation}
where $\text{Pre}_{i}$ denotes the predicted unique author that paper $i$ belongs to. Then, we compare the predicted matrix $\textbf{Pre}$ with the ground-truth matrix $\textbf{L}$ to dynamically identify simple and hard samples.

{\textbf{Simple samples:} Paper pairs that are correctly clustered by HAC ($\textbf{Pre}_{ij}=\textbf{L}_{ij}$). These pairs are mostly from prolific authors with consistent research topics and collaboration patterns, which are easy for the model to learn but mask the discriminative features of tail authors.}

{\textbf{Hard samples:} Paper pairs that are incorrectly clustered by HAC ($\textbf{Pre}_{ij}\neq\textbf{L}_{ij}$). These pairs are mostly from tail authors with limited training signals, which contain critical information for learning clear boundaries between different authors but are overlooked by traditional contrastive learning.}

Based on this dynamic sample division, we reweight the BCE loss to reduce the influence of simple samples and amplify the influence of hard samples, forcing the model to learn more comprehensive and balanced discriminative features. The final harmony contrastive learning loss is formulated as:
\begin{equation}
\label{DCLequation}
  \mathcal{ L}^{HCL} = \alpha \times BCE^{Simple}+(1-\alpha)\times
 BCE^{Hard},
\end{equation}
where $BCE^{Simple}$ is the BCE loss of simple samples, $BCE^{Hard}$ is the BCE loss of hard samples, and the weight parameter $\alpha$ is set to be smaller than 0.5 (empirically 0.2 in our experiments) to ensure that hard samples dominate the training signal. The division of simple and hard samples changes dynamically during training: as the model learns better embeddings, more tail author pairs are correctly clustered and become simple samples, which means the model gradually masters the discriminative features of all authors.

{In summary, HCL directly addresses the core defect of Limitation 1 from the representation learning perspective: it suppresses the model's overfitting to prolific authors by dynamically reweighting the training loss, and learns a high-discriminative embedding space with clear boundaries between both prolific and tail authors. This high-quality embedding space provides a solid foundation for the subsequent GRDM module, which further eliminates the remaining prolific bias from the clustering perspective.}

\subsection{Graph-Refined Distance Matrix (GRDM)}
{Although the proposed HCL module effectively alleviates the model's bias towards prolific authors and learns a more discriminative embedding space, we empirically find that directly applying conventional distance-based clustering (e.g., vanilla HAC) on the learned embeddings still suffers from the over-merging of less-published tail authors into large prolific clusters. The root reason is two-fold: (1) The initial heterogeneous graph topology is inherently biased towards prolific authors, who have far more co-author, co-organization, and co-venue connections, thus dominating the graph structure and misleading the clustering process; (2) Graph Neural Networks (GNNs) inherently rely on message passing mechanisms to aggregate features from neighboring nodes, which tends to make embeddings of adjacent nodes more similar in the feature space, a phenomenon known as feature homogenization or over-smoothing~\cite{li2018deeper}. In our heterogeneous paper graph, papers from prolific authors have much higher node degrees and more diverse neighbors, which further amplifies this homogenization effect: tail authors' papers with few connections are easily "assimilated" by the embeddings of adjacent prolific papers, even after HCL improves the overall discriminability, leading to over-merging in the final clustering stage.}

{To address the above issues and completely solve Limitation 1 from the clustering perspective (complementary to HCL which solves it from the representation learning perspective), we propose the Graph-Refined Distance Matrix (GRDM) module, which directly optimizes the clustering distance matrix based on the learned high-quality embeddings from HCL, without introducing additional training objectives or trade-offs. GRDM consists of two sequential steps: topology refinement and degree-aware similarity adjustment, both explicitly designed to prevent the over-merging of tail authors.}

\subsubsection{Topology Refinement}
We first correct the initial biased heterogeneous graph topology based on the learned paper embedding similarity matrix $\mathbf{S}$ (from HCL), which eliminates the inherent dominance of prolific authors in the initial graph structure. Formally, given the initial adjacency matrix $\mathbf{A}$ (constructed from co-author, co-organization, and co-venue relations) and the embedding similarity matrix $\mathbf{S}$, we refine the topology as:
\begin{equation}
\label{refine}
\mathbf{A'} =
\begin{cases}
1 & \text{if } S_{ij} > \gamma \\
0 & \text{if } S_{ij} < \psi\\
A_{ij} & \text{others} ,
\end{cases}
\end{equation}
where $\mathbf{A'}$ is the refined adjacency matrix, $\gamma$ and $\psi$ are pre-defined thresholds for adding and removing edges, respectively. This step removes noisy edges that are dominated by prolific authors but have low semantic similarity, and adds reliable edges between tail authors with high semantic similarity, thus correcting the initial topology bias.

\subsubsection{Degree-Aware Similarity Adjustment}
On top of the refined topology, we further adjust the similarity matrix to explicitly leverage the node degree information, which is a strong indicator for distinguishing prolific authors and tail authors. Intuitively, a paper node with a higher degree (usually from a prolific author) has more credible co-author, co-organization, and co-venue connections, thus its embedding is more informative and stable. Conversely, a paper node with a lower degree (usually from a tail author) has fewer connections, and its embedding is more likely to be affected by noise. Conventional HAC only focuses on pairwise embedding similarity, which often leads to the over-merging of tail authors into prolific clusters due to accidental semantic similarity. To address this, we first convert the degree of each paper node into a confidence score of its embedding reliability:
\begin{equation}
\label{level}
Score_{i} = 2-\frac{1}{d_{i}+1},
\end{equation}
where $d_{i}$ is the degree of paper node $i$ in the refined adjacency matrix $\mathbf{A'}$. The score ranges from 1 to 2, where higher degrees correspond to higher confidence.

Then, we use this confidence score to adaptively adjust the similarity matrix $\mathbf{S}$, with the explicit goal of preventing the over-merging of tail authors:
\begin{equation}
\label{graph enhance}
\mathbf{S'} =
\begin{cases}
\frac{S_{ij}(Score_{i}+Score_{j})}{2}&  \text{if } A_{ij} ==1 \\
S_{ij} & \text{others}, \\
\end{cases}
\end{equation}
Equation~(\ref{graph enhance}) adjusts the pairwise similarity based on both the refined topology and node degrees: for connected paper pairs (credible relations), we amplify the similarity of pairs with high average confidence (both from prolific authors or both from tail authors with stable connections), while relatively reducing the similarity of pairs with mismatched confidence (one from a prolific author and one from a tail author). This mechanism directly avoids that tail authors are wrongly merged into prolific clusters due to accidental semantic similarity. On the other hand, for unconnected pairs, we keep the original similarity to preserve the global semantic structure.

Finally, we convert the adjusted similarity matrix $\mathbf{S'}$ into the final graph-refined distance matrix $\mathbf{D}$ for HAC clustering:
\begin{equation}
\label{matrix}
\mathbf{D}=S'_{max}-\mathbf{S'},
\end{equation}
where $S'_{max}$ is the maximum similarity value in $\mathbf{S'}$.

{In summary, GRDM works in synergy with HCL to completely solve Limitation 1: HCL alleviates the prolific bias from the representation learning perspective by adaptively weighting the training loss, while GRDM eliminates the remaining bias from the clustering perspective by refining the topology and adjusting the distance matrix based on node degrees, ensuring that tail authors are not over-merged into prolific clusters.}

\subsection{Cluster Number Estimation with Paper Compression Module (PCM)}
{As identified in Limitation 2, reliable cluster number ($\mathcal{K}$) estimation is an indispensable prerequisite for accurate end-to-end name disambiguation, yet existing methods suffer from severe scalability issues and unreliable performance as the number of papers grows. }

Most previous SND works either require $\mathcal{K}$ as a pre-specified parameter (unrealistic in real-world scenarios) or adopt heuristic-based or neural network-based estimation methods with inherent flaws: (1) Heuristic-based methods like X-means~\cite{pelleg2000extending} and optimal modularity partitioning~\cite{qiao2019unsupervised} iteratively search for optimal $\mathcal{K}$ with predefined criterion functions, which are computationally inefficient and tend to over-merge clusters when dealing with large-scale uneven data; (2) DBSCAN-based methods without explicit $\mathcal{K}$ input are extremely sensitive to hand-crafted distance thresholds, failing to generalize across different datasets; (3) Existing neural network-based methods~\cite{zhang2018name} achieve modest performance but rely on fixed-length paper sequences and standardized category ranges, which cannot handle the long and uncertain paper sequences in real academic data due to the inherent long-sequence modeling limitation of RNN architectures (e.g., gradient vanishing/exploding).

{To address the above issues and completely solve Limitation 2, we propose the Paper Compression Module (PCM), a lightweight yet effective neural network-based cluster number estimator that achieves accurate and robust $\mathcal{K}$ prediction across different paper scales. The core insight is to decouple the long-sequence modeling problem into two steps: first compressing the extensive paper embeddings into a compact fixed-length sequence via cross-attention, then processing the compressed sequence with Bi-LSTM for stable and accurate prediction, which eliminates the long-sequence limitation of traditional RNN-based methods.}

Specifically, given a set of paper embeddings $\mathbf{P_{1}}, \mathbf{P_{2}}, \dots, \mathbf{P_{n}}$ (where $n$ can be extremely large and varies across different ambiguous names), we first introduce a set of fixed-length learnable parameters $\mathbf{Q_{n}}$ as queries, and feed them into a cross-attention module with the paper embeddings as both keys and values. This cross-attention mechanism automatically extracts and compresses the most critical information related to cluster number estimation from the extensive paper embeddings into the compact $\mathbf{Q_{n}}$:
\begin{equation}
\label{cross attention}
\mathbf{Q_{n}}=CrossAttention(\mathbf{Q_{n}},(\mathbf{P_{1}},\mathbf{P_{1}}),\dots,(\mathbf{P_{n}},\mathbf{P_{n}})).
\end{equation}

Next, the compressed fixed-length sequence $\mathbf{Q_{n}}$ is fed into a Bi-LSTM to capture both forward and backward contextual information. We take the sum of the forward last hidden state $\mathbf{H^{F}_{n}}$ and the backward last hidden state $\mathbf{H^{B}_{n}}$ as the final compressed paper representation. In addition to the paper embeddings, we also incorporate two types of useful structural auxiliary information: the number of subgraphs $S_{num}$ and the total number of papers $P_{num}$. These two scalars are projected into high-dimensional vectors $\mathbf{SN}$ and $\mathbf{PN}$ via two independent two-layer MLPs, respectively. Finally, we concatenate $\mathbf{SN}$, $\mathbf{PN}$ and the sum of $\mathbf{H^{F}_{n}}$ and $\mathbf{H^{B}_{n}}$, and feed the concatenated vector into a MLP decoder to predict the final cluster size $\mathcal{K}$:
\begin{equation}
\label{cluster size out}
\mathcal{K}=MLP([\mathbf{PN;SN;\mathbf{H^{F}_{n}}+\mathbf{H^{B}_{n}}}]).
\end{equation}

Since the number of author categories varies over a large range in real data, traditional methods~\cite{zhang2018name,qiao2019unsupervised} often use Mean Squared Logarithmic Error (MSLE) as the loss function to ensure numerical stability. However, MSLE limits the model's ability to learn with high precision, especially for small cluster sizes. In contrast, Huber Loss combines the advantages of MSE (Mean Squared Error) and MAE (Mean Absolute Error): when the absolute error between the predicted and true values is within a threshold $\delta$, it employs MSE for high-precision fitting; otherwise, it uses MAE to limit the penalty to linear growth as the error increases, which reduces the impact of outliers and enhances overall prediction accuracy. As a result, we calculate the Huber Loss $L_{\delta}(TN, \mathcal{K})$ between the true cluster size $TN$ and the model output $\mathcal{K}$:
\begin{equation}
\label{MSLE}
L_{\delta}(TN, \mathcal{K}) = \sum_{t=1}^{N}
\begin{cases}
\frac{\frac{1}{2}(TN_{t} - \mathcal{K}_{t})^2}{N} & \text{if } |TN_{t} - \mathcal{K}_{t}| \leq \delta, \\
\frac{\delta |TN_{t} - \mathcal{K}_{t}| - \frac{1}{2}\delta^2}{N} & \text{else}.
\end{cases}
\end{equation}

Finally, Hierarchical Agglomerative Clustering (HAC) with the average linkage criterion takes the predicted cluster size $\mathcal{K}$ (from PCM) and the graph-refined distance matrix $\mathbf{D}$ (from GRDM) as inputs, and outputs the final author clusters.

\subsection{Algorithmic Workflow of GRAND-HC}

\IncMargin{1em}
\begin{algorithm}
\LinesNumbered
\SetKwInOut{Input}{Input}
\SetKwInOut{Output}{Output}
\caption{The Proposed Framework: GRAND-HC}\label{alg}  
\Input{$\mathcal{P}^{na} = \{p_1^{na}, \dots, p_N^{na}\}$}
\Output{$\mathcal{C}^{na} = \{\mathcal{C}_1^{na}, \mathcal{C}_2^{na}, \dots, \mathcal{C}_K^{na}\} $}
\BlankLine
Construct paper relational heterogeneous graph $\mathbf{G}$\;
Compute initial paper feature $\mathbf{X}=[\mathbf{X}^{0},\mathbf{X}^{1},\mathbf{X}^{2},\mathbf{X}^{3}]$, where $\mathbf{X}^{0}\leftarrow BERT(\mathcal{P}^{na})$, $\mathbf{X}^{1}\leftarrow \text{subgraph label}$, $\mathbf{X}^{2}\leftarrow \text{degree}$ and $\mathbf{X}^{3}\leftarrow \text{year}$\;
\For{$iter =1,2,\dots,T$}{ 
    Compute paper embedding:\ $\mathbf{H} \leftarrow \text{Paper Embedding Generation}(\mathbf{X,G})$\;
    Compute harmony contrastive loss:\ $\mathcal{L}^{HCL} \leftarrow \alpha \times BCE^{\text{Simple}}+(1-\alpha)\times BCE^{\text{Hard}}$\;
    Update network parameters\;
}
Optimize graph structure according to (\ref{refine})\;
Compute node confidence score as (\ref{level})\;
Compute refined distance matrix $\mathbf{D}$ according to (\ref{graph enhance})-(\ref{matrix})\;
\For{$iter =1,2,\dots,T$}{ 
    Estimate cluster size:\ $\mathcal{K} \leftarrow PCM(\mathbf{H},S_{num},P_{num})$\;
    Compute Huber loss:\ $L_{\delta}(TN, \mathcal{K})$\;
    Update PCM parameters\;
}
Cluster papers: $\mathcal{C}^{na}\leftarrow HAC(\mathcal{K},\mathbf{D })$
\end{algorithm}
\DecMargin{1em}

The whole process of GRAND-HC is summarized in Algorithm~\ref{alg}. The time complexity for the paper embedding generation network, which utilizes a transformer encoder and graph attention network, is given by $\mathcal{O}(16 \cdot d + N \cdot D \cdot d)$, where $d$ represents the embedding size, $N$ denotes the number of nodes, and $D$ is the average number of neighbors for each node in the graph. The cost of harmony contrastive learning mainly depends on HAC, so it is $\mathcal{O}(N^{2} \cdot d + N^{2} \cdot \log(N) + TN^{3} \cdot d)$, where $TN$ indicates the number of clusters. Besides, cluster size estimation network costs $\mathcal{O}(S \cdot N \cdot d + S \cdot (d \cdot h + h^{2}))$, with $S$ representing the length of the query and $h$ being the hidden embedding size in the Bi-LSTM. As for the graph-refined distance matrix, the cost is solely dependent on the computation of the distance matrix and can be calculated as $\mathcal{O}(N^{2} \cdot d)$. As a result, the whole time complexity of our frame work is $\mathcal{O}((N \cdot D +N^{2}+TN^{3}+S \cdot N) \cdot d+N^{2} \cdot \log(N))$.

\section{Experiments}
\label{sec:experiments}
\subsection{Experiment Setup}
\subsubsection{Datasets}

We use two commonly used author name disambiguation datasets, AMiner-v2 and WhoisWho-v1. AMiner-v2 contains 400 training names,100 validating names, 100 testing names, while WhoisWho-v1 contains 221 training names, 50 validating names, 50 testing names. The details are shown in Table~\ref{dataset}. It can be seen that the distribution of the number of authors per name and the number of papers per name is extremely uneven. For example, the maximum papers per name of WhoisWho-v1 dataset reaches 5682, far exceeding the average of 911.1.

\begin{table}
\small\sf\centering
\caption{The statistics of AMiner-v2 and WhoisWho-v1 datasets.}
\label{dataset}
\begin{tabular}{p{3.5cm}cc}
\toprule
\textbf{Statictics}&\textbf{AMiner-v2}&\textbf{WhoisWho-v1} \\
\midrule
total names & 600 & 321\\
total authors&39,781&26,093\\
total papers&208,827&292,488\\ \midrule
\# of authors/name &2/542/66.3&0/588/81.3\\ \midrule
\# of papers/name &192/916/348.0&0/5682/911.1\\
\bottomrule
\end{tabular}
\end{table}

\subsubsection{Evaluation Metrics}

To evaluate the model performance, we use pairwise precision, pairwise recall and pairwise F1-score as evaluation metrics for each name:
\begin{equation}
Pre=\frac{\#PC}{\#TPP},\quad Rec=\frac{\#PC}{\#TP},\quad F1=\frac{2 \times Pre \times Rec}{Pre + Rec},
\end{equation}
where $\#TPP$ means the number of total pairs predicted to the same author, $\#TP$ means the number of total pairs belongs to the same author and $\#PC$ means pairs correctly predicted pairs to the same author. Here, we use macro precision, recall and F1 as final metric:
\begin{align}
\text{Macro Pre} &= \frac{1}{n} \sum_{i = 1}^{n} Pre_i = \frac{1}{n} \sum_{i = 1}^{n} \frac{\#PC_i}{\#TPP_i},\\
\text{Macro Rec} &= \frac{1}{n} \sum_{i = 1}^{n} Rec_i = \frac{1}{n} \sum_{i = 1}^{n} \frac{\#PC_i}{\#TP_i},\\
\text{Macro F1} &= \frac{2 \times \text{Macro Pre} \times \text{Macro Rec}}{\text{Macro Pre} + \text{Macro Rec}},
\end{align}
where n is the total number of name in the dataset.

\subsubsection{Implementation Details}
We use PyTorch to implement the scheme, the optimizer uses AdamW~\cite{loshchilov2017decoupled}, the learning rate is chosen from 0.01 to 0.00001, the number of training epoch is 200, the dropout ratio is 0.5. Besides, we utilize pre-trained OAG-BERT-V2~\cite{xiao2021oag} to get initial paper embeddings.

For model training, we use Optuna to obtain the best hyper parameters, which include the final paper embedding size from 64 to 128, the thresholds $\psi$ from 0.05 to 0.5 and $\gamma$ from 0.5 to 0.95, and the value of $\alpha$ in the harmony contrastive learning selected from 0 to 0.5. The threshold $\delta$ in Huber loss is selected as 20. The sequence length of query in PCM is 24. The initial graph of AMiner-v2 is constructed by three co-authors and WhoisWho-v1 is two co-authors and one co-organization. All experiments are conducted on one RTX3090.

\subsection{Overall Results}

\begin{table}
\small\sf\centering
\setlength{\tabcolsep}{3pt}
\caption{The performance comparison of AMiner-v2.}
\label{aminer_result}
\begin{tabular}{l|ccc|c}
\toprule
\textbf{Model} & \multicolumn{3}{c|}{\textbf{AMiner-v2}} & \textbf{T(s)} \\
\cmidrule{2-4}
& \textbf{Pre} & \textbf{Rec} & \textbf{F1} & \\
\midrule
Beard & 57.09 & 77.22 & 63.10 & 82.5 \\
AGAND & 70.63 & 59.53 & 62.81 & 78.3 \\
AMiner & 77.96 & 63.03 & 67.79 & 95.2 \\
\underline{ITAND} & 78.10 & 67.47 & 72.40 & 88.6 \\
MFAND & 81.39 & 69.47 & 74.92 & 76.4 \\
MGATAND & 83.87 & 64.91 & 73.10 & 99.7 \\
\underline{OAG-BERT-V2} & 74.26 & 50.15 & 56.41 & 105.3 \\
\underline{MORE} & 74.09 & 77.95 & 76.21 & 112.8 \\
MRAND & 72.40 & 75.10 & 71.50 & 89.4 \\
ARCC & 78.09 & 82.32 & 78.96 & 125.6 \\
\underline{BOND} & 76.55 & 63.40 & 66.21 & 108.2 \\
GPT-2 & 45.20 & 52.30 & 48.50 & 182.4 \\
{Qwen3-4B} & {69.82} & {72.53} & {70.11} & {157.9} \\
{GRAND} & {82.39} & {84.61} & {80.76} & {65.7} \\
\textbf{GRAND-HC(our)} & $\mathbf{88.07_{\pm 0.84}}$ & $\mathbf{88.37_{\pm 0.67}}$ & $\mathbf{84.86_{\pm 0.54}}$ & \textbf{62.9} \\
\bottomrule
\end{tabular}
\end{table}

\begin{table}
\small\sf\centering
\setlength{\tabcolsep}{3pt}
\caption{The performance comparison of WhoisWho-v1.}
\label{who_result}
\begin{tabular}{l|ccc|c}
\toprule
\textbf{Model} & \multicolumn{3}{c|}{\textbf{WhoisWho-v1}} & \textbf{T(s)} \\
\cmidrule{2-4}
& \textbf{Pre} & \textbf{Rec} & \textbf{F1} & \\
\midrule
Beard & 72.20 & 46.19 & 56.34 & 85.3 \\
AGAND & 76.40 & 35.20 & 48.19 & 79.8 \\
AMiner & 77.70 & 55.50 & 64.75 & 96.7 \\
\underline{ITAND} & 59.47 & 65.80 & 61.31 & 90.2 \\
MFAND & 73.36 & 81.03 & 77.00 & 78.5 \\
MGATAND & 67.98 & 79.99 & 73.45 & 101.3 \\
\underline{OAG-BERT-V2} & 76.61 & 84.07 & 78.98 & 108.4 \\
\underline{MORE} & 78.48 & 84.47 & 81.24 & 115.6 \\
ARCC & 79.52 & 83.15 & 80.11 & 128.3 \\
\underline{BOND} & 78.52 & 91.07 & 83.13 & 110.5 \\
GPT-2 & 42.80 & 48.60 & 45.50 & 176.9 \\
{Qwen3-4B} & {59.72} & {66.84} & {63.19} & {168.2} \\
{GRAND} & {75.49} & {76.33} & {77.80} & {71.5} \\
\textbf{GRAND-HC(our)} & $\mathbf{81.06_{\pm 0.69}}$ & $\mathbf{91.23_{\pm 0.36}}$ & $\mathbf{84.33_{\pm 0.58}}$ & \textbf{64.5} \\
\bottomrule
\end{tabular}
\end{table}

Since our architecture consists of paper embedding generation and cluster size estimation modules, we design \textbf{two unified evaluation protocols} for fair comparison, which completely eliminates the setting inconsistency. For a fair comparison on the core disambiguation ability, {\textbf{Protocol 1 (fixed cluster size)} specifies the ground-truth number of authors for all methods, and we remove our PCM module to ensure consistency. Under this protocol, we compare with feature-based methods~\cite{louppe2016ethnicity, zhou2021multiple, xiao2021oag}, graph-based methods~\cite{zhang2017name, ma2020graph, zhang2021author, pooja2022exploiting, gong2024more, liu2024author, cheng2024bond, huang2025framework}, and LLM-based methods~\cite{radford2019language, yang2025qwen3}, so as to verify the effectiveness of our HCL and GRDM. To validate the full pipeline with cluster size estimation, \textbf{Protocol 2 (automatic cluster number)} only selects baselines that can predict the cluster number by themselves, and all methods use their own estimation without ground-truth supervision.} Note that results with underlines are reproduced by our implementation, others are from the original papers.
\begin{figure*}[h]
  \centering
  \includegraphics[width=\textwidth]{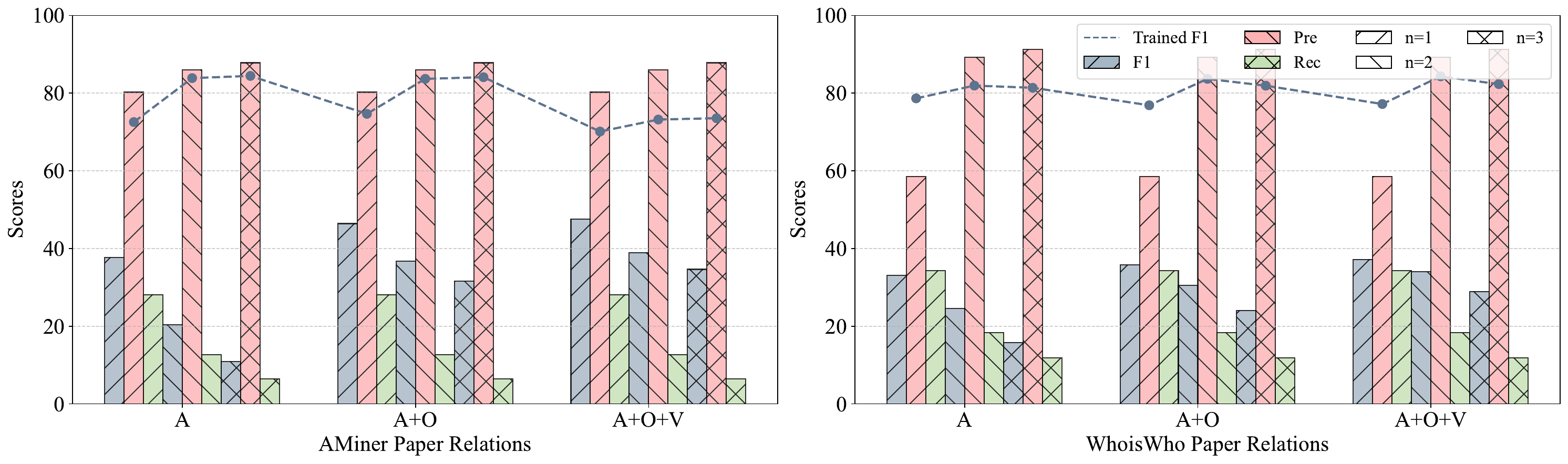}
  \caption{F1, precision and recall score of different paper relational graphs and their final trained F1. A: CoAuthor, O: CoOrg, V: CoVenue. For each paper relation, the co-authorship is analyzed under three scenarios: at least one, two, and three co-authors, as shown by n in the legend. Trained F1 represents the final training results using the corresponding construction principles.}
  \label{fig:chushai}
\end{figure*}

\begin{itemize}
\item \textbf{Beard 2016~\cite{louppe2016ethnicity}:} Beard introduces a novel automated disambiguation solution leveraging over one million crowdsourced annotations, enhancing state-of-the-art methods through phonetic-based blocking strategies, ethnicity-sensitive features, and balanced training for improved accuracy in author name disambiguation.
\item \textbf{AGAND 2017~\cite{zhang2017name}:} AGAND proposes a novel SND method that leverages anonymized graph data and representation learning to partition documents into unique individuals, outperforming existing approaches in similar settings.
\item \textbf{AMiner 2018~\cite{zhang2018name}:} AMiner presents a representation learning method that combines global and local information and demonstrates an end-to-end cluster size estimation method.
\item \textbf{ITAND 2020~\cite{ma2020graph}:} ITAND combines a Variational Graph Auto-Encoder (GAE) with graph embedding to capture attribute and topological features, employing hierarchical agglomerative clustering for document partitioning.
\item \textbf{MFAND 2021~\cite{zhou2021multiple}:} MFAND refines and merges the different raw graph information, and then uses the convolution-based R3JG encoder to learn the relationships between text pairs directly, thus avoiding clustering algorithms.
\item \textbf{MGATAND 2021~\cite{zhang2021author}:} MGATAND reconstructs heterogeneous graphs into homogeneous graphs, then uses the reduced topology as the training goal of the GAT network. Finally, spectral clustering algorithm is employed without specific cluster size.
\item \textbf{OAG-BERT-V2 2021~\cite{xiao2021oag}:} OAG-BERT-V2 is a BERT model pretrained on the academic text corpus in Open Academic Graph, including paper titles, abstracts and bodies and finetuned on several SND datasets.
\item \textbf{MRAND 2022~\cite{pooja2022exploiting}:} MRAND proposes a multidimensional multi-hop neighbor graph convolutional network based on an attention mechanism, which effectively utilizes the multi-order neighbor information of different relationship types in heterogeneous graphs.
 \item \textbf{MORE 2024~\cite{gong2024more}:} MORE integrates many advanced optimization and SND techniques such as OAG-BERT, SimCSE, LightGBM, and iHGAT via supervised and unsupervised learning.
\item \textbf{ARCC 2024~\cite{liu2024author}:} ARCC employs an iterative process to refine the graph structure of paper graphs, dynamically reducing uncertainties. It trains its model using contrastive learning and uses HAC for clustering.
\item \textbf{BOND 2024~\cite{cheng2024bond}:} BOND uses DBSCAN predictions to construct pseudo-labels during training, exploiting local pairwise similarity to drive global clustering. Finally, its clusters by the DBSCAN algorithm with fine-tuned thresholds.

\item {\textbf{GRAND 2025~\cite{huang2025framework}:} GRAND is a global role-based author name disambiguation framework, which adopts meta-path guided embedding and solid co-author sampling to address ambiguous co-authorship. It distinguishes real-world researchers from their author roles, and uses DBSCAN to perform adaptive clustering without manually specifying the cluster number.}

\item \textbf{GPT-2 2019~\cite{radford2019language}:} GPT-2 directly employs the pre-trained large language model for author name disambiguation by encoding paper textual attributes (titles, abstracts, and venues) into dense semantic embeddings. It performs zero-shot clustering based on transformer-based text similarity without leveraging citation graphs or co-author networks, serving as an early-stage LLM baseline.

\item {\textbf{Qwen3-4B 2025~\cite{yang2025qwen3}:} Qwen3-4B is a modern lightweight open-source large language model, which encodes paper textual attributes (titles, abstracts, and venues) into dense semantic embeddings for zero-shot clustering. It performs disambiguation purely based on semantic similarity without using citation graphs or co-author networks, serving as a strong representative of contemporary LLMs to complement the outdated GPT-2 baseline.}
\end{itemize}

\begin{table}[t]
\small
\centering
\setlength{\tabcolsep}{2pt}
\caption{{Performance comparison under unspecified cluster size (vs. fixed cluster size setting).}}
\label{unspecified_k_result}
\begin{tabular}{l|cc|cc} 
\hline
{\textbf{Model}} & \multicolumn{2}{c|}{{\textbf{F1 Score}}} & {\textbf{RMLSE}} & {\textbf{T(s)}} \\
\cline{2-3} 
& {\textbf{AMiner-v2}} & {\textbf{WhoisWho-v1}} & & \\
\hline
{AMiner} & {62.1($\downarrow$5.7)} & {54.2($\downarrow$10.6)} & {0.25} & {103.2($\uparrow$7.3)} \\
{BOND} & {57.3($\downarrow$8.9)} & {74.5($\downarrow$8.6)} & {0.77} & {121.8($\uparrow$12.5)} \\
{GRAND} & {78.5($\downarrow$2.3)} & {67.1($\downarrow$10.7)} & {0.52} & {73.8($\uparrow$5.2)} \\
{\textbf{GRAND-HC}} & {\textbf{84.9($\downarrow$0.0)}} & {\textbf{84.2($\downarrow$0.1)}} & {\textbf{0.18}} & {\textbf{67.4($\uparrow$3.7)}} \\
\hline
\end{tabular}
\end{table}
From the Table~\ref{aminer_result} and Table~\ref{who_result}, it can be seen that our model outperforms most of the current state-of-the-art methods in terms of macro F1 score.

\begin{table*}[ht]
\small\sf\centering
\caption{The detailed results on AMiner.}
\label{aminer detail}
\tabcolsep 0.03in
\begin{tabular}{c|ccc|ccc|ccc|ccc|ccc}
\toprule
\textbf{Name} & \multicolumn{3}{c|}{\textbf{GRAND-HC}} & \multicolumn{3}{c|}{\textbf{BOND}} & \multicolumn{3}{c|}{\textbf{ARCC}} & \multicolumn{3}{c|}{\textbf{MFAND}} & \multicolumn{3}{c}{\textbf{AMiner}}  \\
\midrule
& \textbf{Pre}&\textbf{Rec}&\textbf{F1}& \textbf{Pre}&\textbf{Rec}&\textbf{F1}& \textbf{Pre}&\textbf{Rec}&\textbf{F1}& \textbf{Pre}&\textbf{Rec}&\textbf{F1}& \textbf{Pre}&\textbf{Rec}&\textbf{F1}\\
\midrule
xu\_xu & \textbf{83.23} & \textbf{85.05} & \textbf{84.13} & 71.32 & 57.08 & 63.41 & 65.15 & 69.25 & 67.14 & 34.59 & 79.44 & 48.20 & 74.18 & 45.86 & 56.68 \\
rong\_yu & 97.13 & 95.41 & 96.26 & 85.00 & 43.30 & 57.37 & \textbf{97.58} & \textbf{97.96} & \textbf{97.77} & 72.31 & 43.83 & 54.58 & 89.13 & 46.51 & 61.12 \\
yong\_tian & \textbf{93.91} & 67.01 & \textbf{78.21} & 84.43 & 52.35 & 64.63 & 64.85 & \textbf{76.68} & 70.27 & 46.12 & 59.42 & 51.93 & 76.32 & 51.95 & 61.82 \\
lu\_han & \textbf{64.07} & \textbf{91.41} & \textbf{75.34} & 48.50 & 27.35 & 34.98 & 53.25 & 46.28 & 49.52 & 37.22 & 51.25 & 43.12 & 51.78 & 28.05 & 36.39 \\
lin\_huang & \textbf{93.89} & \textbf{79.72} & \textbf{86.22} & 85.99 & 46.20 & 60.10 & 82.98 & 59.02 & 68.98 & 60.80 & 52.40 & 56.29 & 77.10 & 32.87 & 46.09 \\
kexin\_xu & 80.77 & 81.49 & 81.13 & 81.05 & 87.52 & 84.16 & \textbf{91.41} & \textbf{98.51} & \textbf{94.83} & 83.50 & 81.93 & 82.71 & 91.37 & 98.64 & 94.87 \\
wei\_quan & \textbf{96.89} & \textbf{96.89} & \textbf{96.89} & 74.41 & 28.57 & 41.29 & 91.03 & 92.61 & 91.81 & 35.72 & 48.67 & 41.20 & 53.88 & 39.02 & 45.26 \\
tao\_deng & \textbf{89.07} & \textbf{91.78} & \textbf{90.40} & 74.96 & 40.58 & 52.65 & 79.54 & 68.23 & 73.45 & 59.55 & 41.22 & 48.72 & 81.63 & 43.62 & 56.86 \\
hongbin\_li & \textbf{94.17} & \textbf{94.67} & \textbf{94.42} & 86.43 & 68.02 & 76.13 & 76.87 & 93.89 & 84.53 & 48.85 & 78.86 & 60.33 & 77.20 & 69.21 & 72.99 \\
hua\_bai & \textbf{98.33} & 57.43 & 72.51 & 73.66 & 38.37 & 52.63 & 90.86 & \textbf{87.04} & \textbf{87.04} & 73.66 & 55.82 & 63.51 & 71.49 & 39.73 & 51.08 \\
mei\_ling\_chen & \textbf{99.06} & \textbf{93.78} & \textbf{96.35} & 82.20 & 38.72 & 52.64 & 58.47 & 85.70 & 69.52 & 94.80 & 41.78 & 58.00 & 74.93 & 44.70 & 55.99 \\
yanqing\_wang & 62.05 & 80.75 & 70.17 & 65.22 & 63.56 & 64.38 & \textbf{88.96} & 46.10 & 46.10 & 70.37 & 57.20 & 63.10 & \textbf{71.52} & 75.33 & \textbf{73.37} \\
xu\_dong\_zhang & \textbf{94.60} & 51.00 & 66.28 & 67.03 & 10.37 & 17.96 & 86.13 & \textbf{59.89} & \textbf{70.66} & 51.48 & 24.17 & 32.90 & 62.40 & 22.54 & 33.12 \\
qiang\_shi & \textbf{71.35} & 66.45 & \textbf{68.81} & 51.55 & 41.13 & 45.76 & 54.46 & 54.42 & 52.90 & 40.53 & \textbf{76.46} & 52.97 & 52.20 & 36.15 & 42.72 \\
min\_zheng & \textbf{87.36} & \textbf{80.27} & \textbf{83.67} & 71.38 & 19.61 & 30.77 & 77.05 & 49.95 & 60.61 & 31.74 & 52.48 & 39.55 & 57.65 & 22.35 & 32.21 \\
\bottomrule
\end{tabular}
\end{table*}

\begin{table*}[ht]
\small\sf\centering
\caption{The detailed results on WhoisWho-v1.}
\label{who detail}
\tabcolsep 0.03in
\begin{tabular}{c|ccc|ccc|ccc|ccc|ccc}
\toprule
\textbf{Name} & \multicolumn{3}{c|}{\textbf{GRAND-HC}} & \multicolumn{3}{c|}{\textbf{BOND}}& \multicolumn{3}{c|}{\textbf{ARCC}} & \multicolumn{3}{c|}{\textbf{MFAND}} & \multicolumn{3}{c}{\textbf{AMiner}}  \\
\midrule
& \textbf{Pre}&\textbf{Rec}&\textbf{F1}& \textbf{Pre}&\textbf{Rec}&\textbf{F1}& \textbf{Pre}&\textbf{Rec}&\textbf{F1}& \textbf{Pre}&\textbf{Rec}&\textbf{F1}& \textbf{Pre}&\textbf{Rec}&\textbf{F1}\\
\midrule
baohong\_zhang &\textbf{100} & \textbf{100} & \textbf{100} & 96.89 & 98.74 & 97.81 & 85.76 & 99.92 & 92.30 & 96.54 & 86.53 & 91.26 & 93.18 & 73.27 & 82.04 \\
aiqin\_wang &\textbf{99.95} & \textbf{99.72} & \textbf{99.84} & 96.28 & 95.41 & 95.84& 98.16 & 96.55 & 97.35 & 99.90 & 88.25 & 93.71 & 90.69 & 99.85 & 95.04 \\
haibo\_he &99.97 & \textbf{98.25} & \textbf{99.10} & 98.23 & 98.24 & 98.23& 99.39 & 93.18 & 96.18 & \textbf{99.99} & 88.14 & 93.69 & 97.92 & 38.53 & 55.30 \\
bing\_ren &\textbf{99.76} & \textbf{99.31} & \textbf{99.54} & 96.76 & 97.54 & 97.15& 96.85 & 97.54 & 97.19 & 91.81 & 72.09 & 80.76 & 90.91 & 97.17 & 93.94\\
jijun\_zhao &\textbf{99.34} & \textbf{98.35} & \textbf{98.84} & 84.41 & 98.35 & 90.84& 86.02 & 91.86 & 88.84 & 98.89 & 84.29 & 91.01 & 94.46 & 95.69 & 95.07 \\
frank\_caruso &72.43 & 96.60 & \textbf{82.79} & 70.21 & \textbf{97.59} & 81.66 & \textbf{81.63} & 71.50 & 76.23 & 79.94 & 60.67 & 68.98 & 76.71 & 36.25 & 49.23 \\
xiaohong\_guan &87.28 & 98.21 & 92.42 & 98.22 & \textbf{99.31} & \textbf{98.76}& \textbf{98.28} & 96.59 & 97.43 & 88.02 & 91.51 & 89.73 & 78.16 & 55.19 & 64.70 \\
david\_parker &62.97 & 97.73 & 76.59 & 63.63 & 97.86 & 77.12& 64.90 & \textbf{98.92} & \textbf{78.38} & \textbf{67.11} & 84.28 & 74.72 & 56.50 & 55.19 & 71.49 \\
hongjun\_song &96.95 & 88.34 & 92.45 & 97.68 & 89.32 & 93.31& 84.54 & 88.44 & 86.45 & \textbf{99.70} & 78.58 & 87.89 & 99.58 & \textbf{95.19} & \textbf{97.34} \\
min\_hu &80.77 & 92.61 & \textbf{86.28} & 75.40 & \textbf{94.50} & 83.87& \textbf{84.10} & 81.95 & 83.01 & 77.54 & 74.94 & 76.22 & 81.27 & 63.93 & 71.57 \\
jie\_tang &85.63 & 94.90 & 90.02 & 83.38 & \textbf{95.62} & 89.08& \textbf{94.98} & 94.90 & \textbf{94.94} & 90.31 & 77.55 & 83.44 & 71.27 & 29.69 & 41.91 \\
feng\_wang &71.97 & \textbf{94.41} & \textbf{81.68} & 70.59 & 93.91 & 80.60 & \textbf{73.71} & 90.15 & 81.11 & 48.02 & 87.22 & 61.94 & 59.71 & 66.67 & 63.00 \\
jian\_pei &\textbf{96.59} & \textbf{99.56} & \textbf{98.06} & 96.32 & 99.52 & 97.89& 96.04 & 96.46 & 96.25 & 59.81 & 95.56 & 73.57 & 96.07 & 65.61 & 77.97 \\
haining\_wang &82.31 & 96.65 & 88.91 & 78.44 & 96.44 & 86.52& 83.28 & \textbf{97.97} & \textbf{90.03} & 54.18 & 81.95 & 65.23 & \textbf{85.82} & 43.13 & 57.41 \\
r\_gupta &87.69 & 82.47 & 85.00 & 86.97 & \textbf{96.30} & \textbf{91.40} & \textbf{90.41} & 86.86 & 88.60 & 83.76 & 58.78 & 69.08 & 92.48 & 79.17 & 85.31 \\
\bottomrule
\end{tabular}
\end{table*}
As shown in the results, MFAND's use of text-pair relations as a training target causes its network to fail to capture global author feature information. The rest of the feature-based methods such as OAG-BERT-V2 and AMiner limit their performance since they do not consider the topology of the graph. As for graph-based methods,{ the recent GRAND neglects the uneven data distribution, resulting in less discriminative embeddings}; ARCC relies on multiple rounds of topology modification, which risks introducing error edges and restricting performance. Meanwhile, BOND suffers from biased pseudo-label propagation based on DBSCAN clustering, while ITAND's GVAE model lacks reliable training objectives for noisy datasets.{ For LLM-based approaches, GPT-2 yields the worst performance (F1 $<$ 50\%), and even the modern Qwen3-4B still falls far short of GRAND-HC due to the lack of task-specific structural inductive biases.} In contrast, our GRAND-HC only refines the graph structure once and fully exploits graph information via the GRDM, achieving superior performance.
\textbf{In terms of computational efficiency, GRAND-HC achieves the fastest inference speed at only 60s per name, significantly outperforming graph-based methods that require 65--130s. This efficiency stems from our single-round graph refinement and lightweight PCM module, avoiding the iterative overhead of ARCC. Moreover, all LLM-based methods incur prohibitive runtime, with GPT-2 consuming a maximum of 3min per name, further validating the superiority of our graph-neural architecture for large-scale academic data processing.}
Overall, the excellent results achieved by GRAND-HC set us apart from all counterparts.

{Under the unknown cluster number setting, most baselines fail to work, and only three valid methods are retained and shown in Table~\ref{unspecified_k_result}: RNN-based AMiner, threshold-dependent density clustering method BOND, and GRAND. For experimental consistency, the RNN module of AMiner and our PCM are trained on the same training set; BOND and GRAND adopt the default clustering thresholds in their official codes. GRAND-HC outperforms all baselines in three aspects. In performance, all baselines suffer significant degradation, while GRAND-HC achieves almost no drop on AMiner-v2 and negligible decline on WhoisWho-v1. In cluster number accuracy, GRAND-HC obtains the lowest RMLSE of 0.18. In time overhead, GRAND-HC brings the smallest extra delay and keeps the fastest inference speed. The results confirm that GRAND-HC maintains stable and superior performance in the realistic unknown cluster number scenario.}

\begin{figure}[htb]
  \centering
  \includegraphics[width=\linewidth,height=6cm]{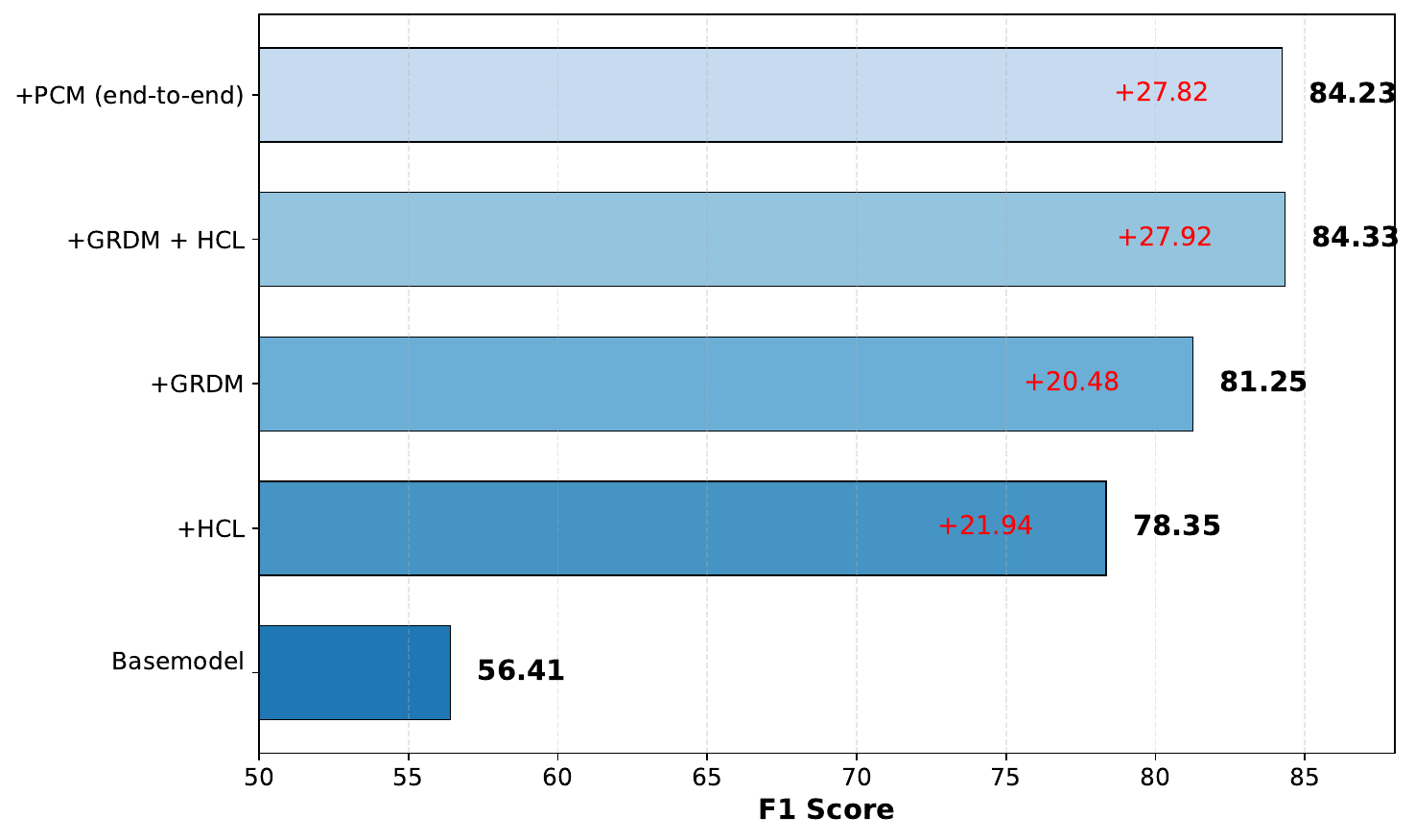}
  \caption{{Ablation study of each key component in GRAND-HC. 
  Baseline denotes our framework equipped with original contrastive learning and HAC clustering. 
  HCL represents harmony contrastive learning. 
  GRDM is the graph-refined distance matrix module. 
  PCM denotes the cluster size estimation module.}}
  \label{fig:ablation}
\end{figure}
{We also conduct an ablation study to verify the contribution of each component (Figure~\ref{fig:ablation}).
The baseline is our framework with standard contrastive learning and HAC clustering, achieving 56.41 F1-score.
By replacing it with harmony contrastive learning (HCL), the performance jumps to 78.35 (+21.94), showing its effectiveness in learning discriminative embeddings.
Adding graph-refined distance matrix (GRDM) further improves the result to 81.25 (+20.48), which validates the role of adaptive distance optimization.
Combining HCL and GRDM yields 84.33 (+27.92), demonstrating their complementary strengths.
Finally, the full model with cluster size estimation (PCM) reaches 84.23, realizing practical end-to-end disambiguation.
These results confirm that each module delivers independent and stable gains, and the improvement comes from our core designs instead of heuristic combinations.}

The results of 15 sampled names are displayed in the Table~\ref{aminer detail} and Table~\ref{who detail}. It can be seen that our model improves on almost all evaluation metrics. However, in the WhoisWho-v1 dataset, there are still some names performance worse than others. We further analyze the classification results for these names relying on the initial graph topology, with an average \textbf{14.78} F1 score(\%). As a result, the graph construction is inaccurate and needs to be iteratively modified like ARCC in order for the network to utilize the topology. But the repeated modifications will increase the computational complexity and reduce the efficiency.

\subsection{Ablation of Paper Relational Heterogeneous Graph Construction}\label{subsec:abla_of_graph}
To fully leverage the diverse relational connections between papers, it is essential to carefully select the types of edges when constructing the heterogeneous graph of academic papers. In this section, we build the heterogeneous graph incrementally based on the Aminer and WhoIsWho training datasets by progressively incorporating different types of edges, aiming to explore their impact on the SND task.

Figure~\ref{fig:chushai} illustrates the F1 performance of different paper relational graphs on the name disambiguation task. Specifically, CoAuthor (A) represents the co-author relationship excluding the target disambiguation author (evaluated under scenarios with at least one, two, and three co-authors), CoOrg (O) represents the co-organization relationship of the author to be disambiguated, and CoVenue (V) represents the co-publication venue relationship between papers. We first construct a heterogeneous graph based on the text-pair relations, and then cluster the articles by dividing the different subgraphs into different authors. Finally, we calculate F1 scores according to the clustered text-pair relations and test their results.

Overall, the AMiner dataset demonstrates superior performance compared to the WhoIsWho dataset. This disparity arises primarily because the WhoIsWho dataset reflects real-world scenarios with substantial noise and missing values. Moreover, considering that co-author relationships excluding the target author may also involve instances of name homonymy, we tested different numbers of co-author connections. The results reveal that the best-performing relational graph in both datasets is A1+O+V, achieving F1 scores of 47.57 and 37.16 for the AMiner and WhoisWho datasets, respectively. This indicates that incorporating multi-relationships effectively mitigates the influence of name homonymy and improves performance.

Furthermore, the performance of A+O shows significant improvement over A alone, highlighting the crucial role of co-organization information in complementing co-author relationships and enhancing name disambiguation. However, co-venue relationships contribute minimally to disambiguation performance. We hypothesize that their primary value lies in the semantic information they convey, such as shared research directions.

\begin{table}
\small\sf\centering
\caption{Graph information influence on F1 score(\%).}
\label{graph_info}
\begin{tabular}{ccc}
\toprule
 &\textbf{AMiner-v2}&\textbf{WhoisWho-v1}\\
\midrule
Initial embedding&56.41&78.98\\
Initial graph& 56.23&50.12 \\
No graph information & 60.58& 82.45\\	
GRAND-HC& \textbf{84.86}&\textbf{84.33}\\
\bottomrule
\end{tabular}
\end{table}

But an interesting phenomenon is that the final training results have a strong correlation with the precision of the initial heterogeneous graph. We think this may be that graph neural networks need more credible connecting edges to aggregate information to generate reliable embedding representations. In addition since we subsequently refine the distance matrix based on the graph structure, topologies with higher initial precision provide more accurate augmentation locations. Overall, GRAND-HC achieves good results on many types of initial heterogeneous graphs with well generalization.

\subsection{Ablation of Paper Embedding Generation}\label{embedding ablation}
This chapter will analyze the importance of each part of our paper embedding generation network.

\begin{figure}[h]
  \centering
  \includegraphics[width=\linewidth,height=3cm]{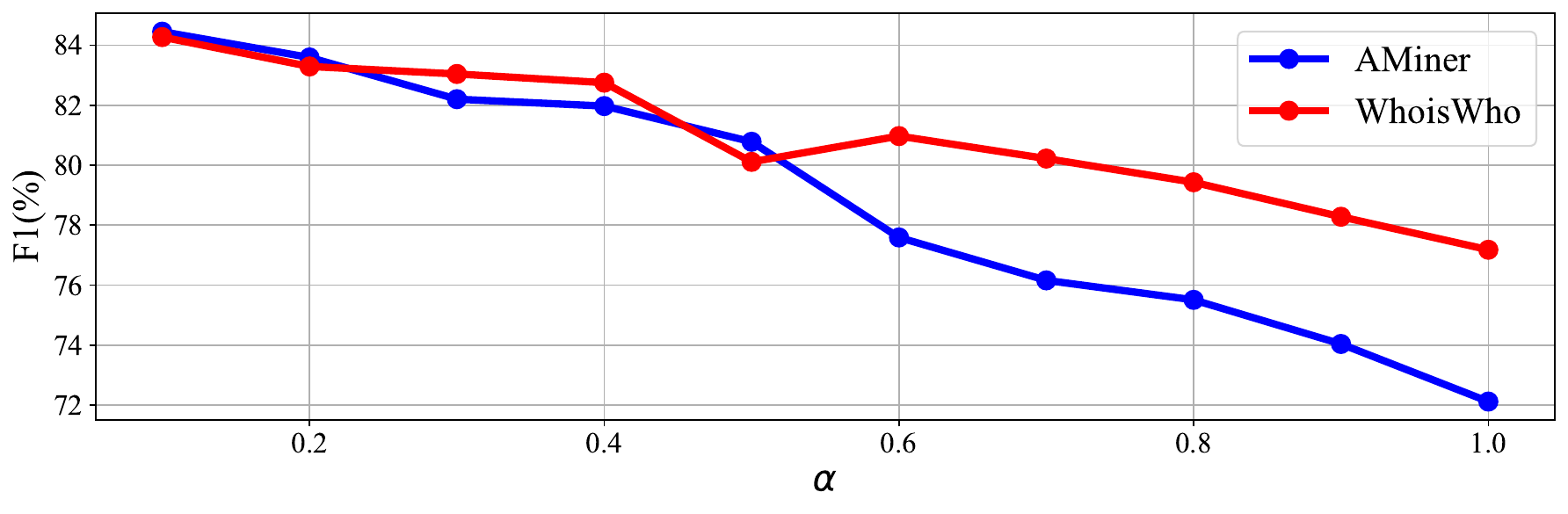}
  \caption{Loss partition parameter $\alpha$ analysis in harmony contrastive learning.}
  \label{a}
\end{figure}

The embedding generation model mainly relies on harmony contrastive learning. We first adjust the loss weight of simple samples $\alpha$ and observe its impact on the final results. In Figure~\ref{a}, it can be seen that the performance of the model decreases significantly as $\alpha$ increases. This indicates that during the training process, most of the samples that can be correctly clustered by HAC come from the papers of high-yield authors with similar features, which are easy to be correctly classified. Overemphasizing the information from these paper pairs can hinder effective feature capture for authors with smaller paper numbers, ultimately resulting in diminished differentiation in paper embeddings. Conversely, moderating the loss contribution of these simple samples during training enables the acquisition of more comprehensive feature representations.

\begin{figure*}[h]
  \centering
  \includegraphics[width=\textwidth]{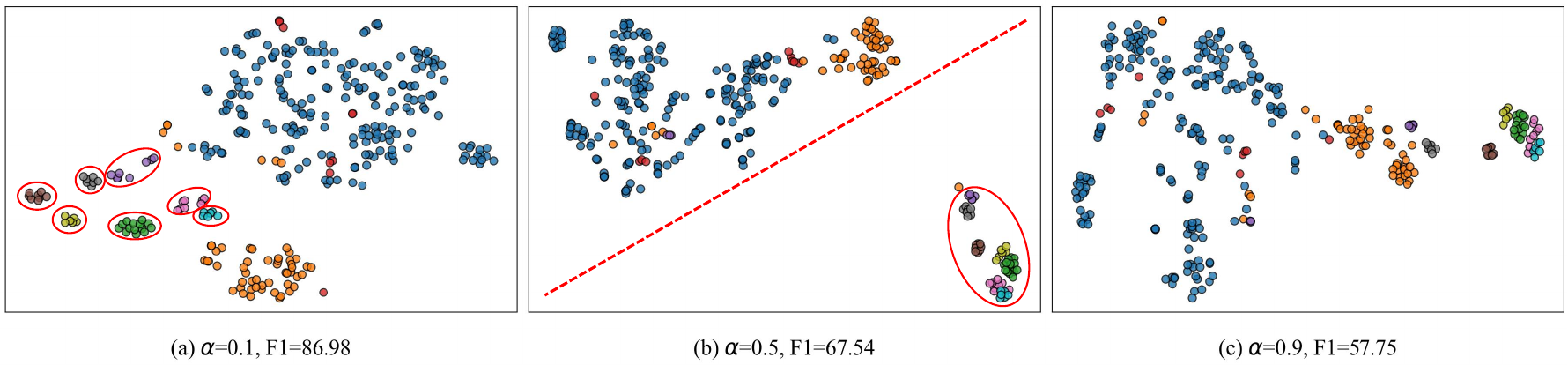}
  \caption{The t-SNE visualization of name \textit{`hongtao\_liu'} in AMiner-v2 for the analysis of parameter $\alpha$.}
  \label{tsne}
\end{figure*}

Specifically, in Figure~\ref{tsne}, we randomly select a name in the AMiner dataset, and use t-SNE to downscale the paper features of the authors of the same name in the top ten according to their paper counts. When $\alpha$ is equal to 0.5, it is equivalent to the traditional contrastive learning loss, i.e., the two kinds of samples have the same percentage. At this time, in the vector space, the model only learns the features of the high-producing authors in blue, and thus the output paper embeddings are mainly distributed in two regions. And when $\alpha$ is equal to 0.9, too many simple samples make the whole vector space more blurred. On the contrary, when we reduce the percentage of these simple samples, we get a more discriminative paper distribution. From the distribution of authors within the red circles in Figure~\ref{tsne}~(a) and~(b), it can be seen that while the high-producing authors represented by the blue dots keep a certain distance from the rest of the authors, the distance between the small authors of the rest of the colors is more significant, which facilitates the accurate clustering of HAC.

In Table~\ref{graph_info}, we conducted ablation experiments on the graph information (subgraph labels and degree) of the model inputs. Initial embedding is clustered directly with the embedding of OAG-BERT-V2, and initial graph is divided directly based on the connected component of the initial structure graph. It can be observed that on the AMiner dataset, the quality of initial semantic embeddings is relatively low, leading to poor performance in direct training. GRAND-HC employs a transformer encoder to jointly encode the structural and semantic information of nodes, which effectively improves the F1-score. Either the initial structural or semantic embeddings can relatively accurately reflect the author clusters, GRAND-HC can learn the underlying patterns with strong generalization.

\subsection{Ablation of Cluster Size Estimation Network}\label{anm}
In order to test the effectiveness of the cluster size estimation network, we selected main solutions for this problem as baselines and use root mean squared logarithmic error as evaluation metric according to previous works. We test them on the AMiner-v2 dataset and sample some author results, which is shown in Table~\ref{author num}. Note that results with underline is reproduced by us and others are from AMiner~\cite{zhang2018name}.

As previously mentioned, existing cluster size estimation algorithms are mainly categorized into four types. We select the SOTA models for each type and carry out a comprehensive comparison. It can be observed that the RMLSE error of our method is far lower than that of the other algorithms.

The X-means~\cite{pelleg2000extending} method relying on the Bayesian Information Criterion cannot handle a complex mixture data in high dimension~\cite{giraud2021introduction}, which deviates greatly from the true value. The graph-based clustering algorithm GHAC~\cite{qiao2019unsupervised} tends to find clusters of moderate size and may fail to accurately identify some extremely small or large clusters. Additionally, potential noisy nodes may disrupt the calculation of edge weights, the measurement of cluster similarity, and the evaluation of modularity. The threshold-based DBSCAN~\cite{cheng2024bond} in BOND is highly parameter-sensitive and prone to over merge clusters due to ineffective distance metrics in high dimensions and noise misinterpretation, often underestimating the true class count. Aminer~\cite{zhang2018name} samples and constructs a standard dataset with a range from 1 to 300 to make it processable for the RNN. Although this achieves a certain generalization effect, in our model, PCM adopts a fixed-length learnable parameter before RNN for information aggregation, avoiding the need to construct a standard dataset and simplifying the processing flow. Meanwhile, Bi-LSTM can process the data more efficiently, reducing the RMLSE metrics by nearly 6\%. We also remove the cross-attention module in PCM, and the RMLSE increases to 0.1940, which demonstrates the necessity of information extraction by the cross-attention module.

\begin{figure}[h]
  \centering
  \includegraphics[width=0.95\linewidth,height=4cm]{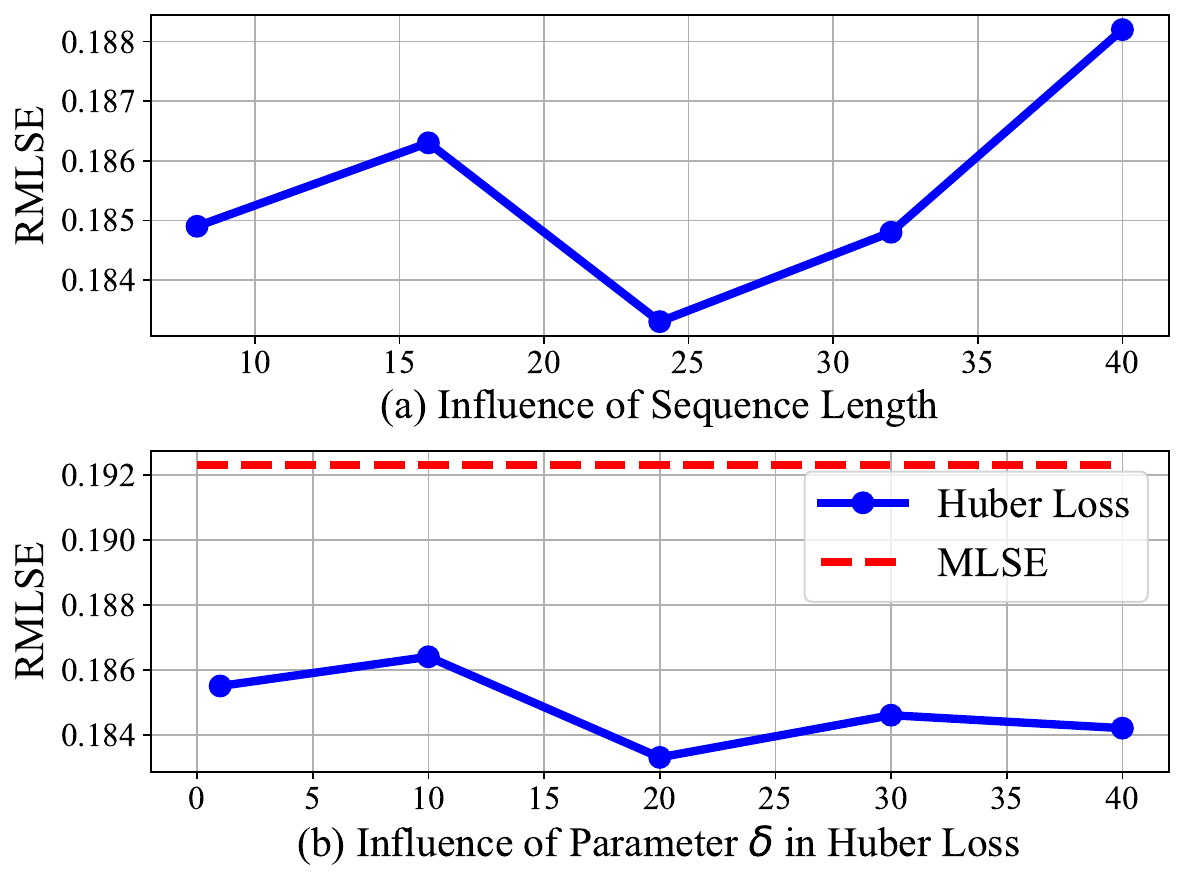}
  \caption{Ablation analysis of PCM. (a) shows the RMLSE changes with the query length. (b) shows the influence of $\delta$ in Huber Loss and difference between MLSE and Huber Loss}
  \label{RMLSE}
\end{figure}

Furthermore, we analyze the effect of sequence length of learnable queries on RMLSE in Figure~\ref{RMLSE}~(a). It can be seen that there exists an optimal sequence length around 25. When the sequence is too long, it may contain excessive repetitive information, while a short sequence may lack sufficient useful data with poorer performance. Additionally, in Figure~\ref{RMLSE}~(b), we discuss the impact of $\delta$ on predicted RMLSE, where $\delta$ controls the switching point between using less punitive MAE and more punitive high-precision MSE. Compared with MLSE, Huber Loss first employs coarse MAE learning and then switches to MSE for fine-grained feature learning when the absolute error between the model's predictions and true values is less than $\delta$, which significantly enhances the model's performance and exists a minimum RMLSE around $\delta$ of 25.

\subsection{Ablation of Graph-refined Distance Matrix}

graph-refined distance matrix(GRDM) consists of two steps: graph structure optimization and graph based distance matrix refinement. We compare our GRDM with another graph-based HAC clustering technique that is similar to our work. In GHAC~\cite{qiao2019unsupervised}, if there is an edge between two papers, their similarity is preserved and normalized by a sigmoid activation function:
\begin{equation}
|S_{ij}| = \sigma(p_{i} \cdot p_{j}) \cdot \delta((p_{i}, p_{j}) \in E),
\end{equation}
where $\sigma(\cdot)$ is the sigmoid function, $\delta(x)$ is $1$ if $x$ is true and $0$ otherwise, $p_{i}$ and $p_{j}$ are paper embeddings, E is the edge set in paper graph G.

\begin{table}
\small\sf\centering
\caption{Comparison of cluster size estimation methods.}
\label{author num}
\tabcolsep 0.02in
\begin{tabular}{ccccccc}
\toprule
 & \textbf{Actual} & \textbf{GRAND-HC} & \textbf{AMiner} & \textbf{\underline{GHAC}} & \textbf{\underline{BOND}} & \textbf{X-means} \\
\midrule
RMLSE & - & \textbf{0.1833}& 0.2493&0.9188&0.7678 & 2.1065 \\
\midrule
Song Chen & 125 & 118 &  101&74& 81& 10 \\
Jian Du & 87 &43 &  63&27&33 & 5 \\
Fosong Wang & 4 & 4&  6&36&10 & 5 \\
J Yu & 346 & 125 &  74&72&71& 7 \\
Yang Shen & 157 & 88 &  154&70&69 & 7 \\
Xiaobing Luo & 13 & 12 &  11&21&11 & 3 \\
Jian Feng & 102 & 58 &  150 &60&51& 8 \\
Lu Han & 129 & 81 &  115&51&56 & 7 \\
\bottomrule
\end{tabular}
\end{table}
\begin{figure*}[h]
  \centering
  \includegraphics[width=\textwidth]{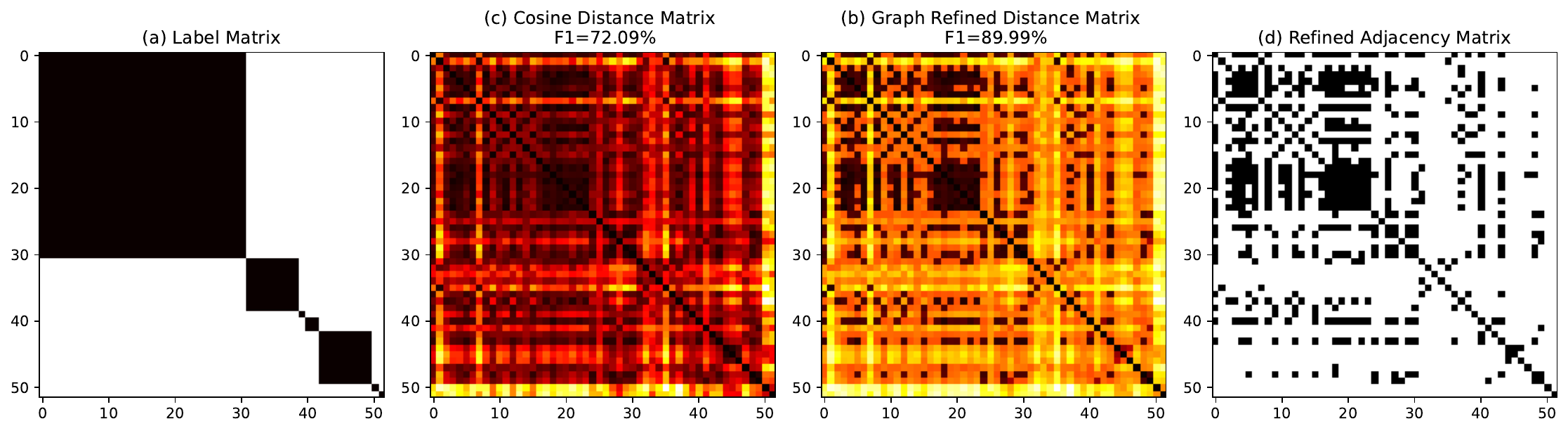}
  \caption{The visualization of name \textit{`philip\_kam\_tao\_li'} in AMiner-v2 for the analysis of graph-refined distance matrix(GRDM). The lighter the color in (a), (b) and (c), the farther the distance.}
  \label{hot}
\end{figure*}
In Table~\ref{graph_pro}, Original represents clustering directly using the cosine distance matrix, while `-o' and `-r' correspond to refine distance matrix based on original graph and optimized graph, respectively. From the results of GHAC-o and GHAC-r, it can be seen that they are extremely dependent on the accuracy of the topology. Whereas GRAND-HC-o, which is adaptively tuned based on degree using~(\ref{graph enhance}), is more robust and GRAND-HC-r has a greater improvement when the structure of the graph is more accurate.

\begin{table}
\small\sf\centering
\caption{Ablation study of graph-refined distance matrix(GRDM) on F1 score (\%).}
\label{graph_pro}
\begin{tabular}{ccc}
\toprule
 &\textbf{AMiner-v2}&\textbf{WhoisWho-v1}\\
\midrule
Original&	81.25	&80.90\\
GHAC-o~\cite{qiao2019unsupervised}& 75.87 &73.45\\
GHAC-r~\cite{qiao2019unsupervised}& 82.47 &81.26\\
GRAND-HC-o	&82.83	&81.98\\
GRAND-HC-r	&\textbf{84.86}	&\textbf{84.33}\\
\bottomrule
\end{tabular}
\end{table}

\begin{figure}[h]
  \centering
  \includegraphics[width=\linewidth,height=4cm]{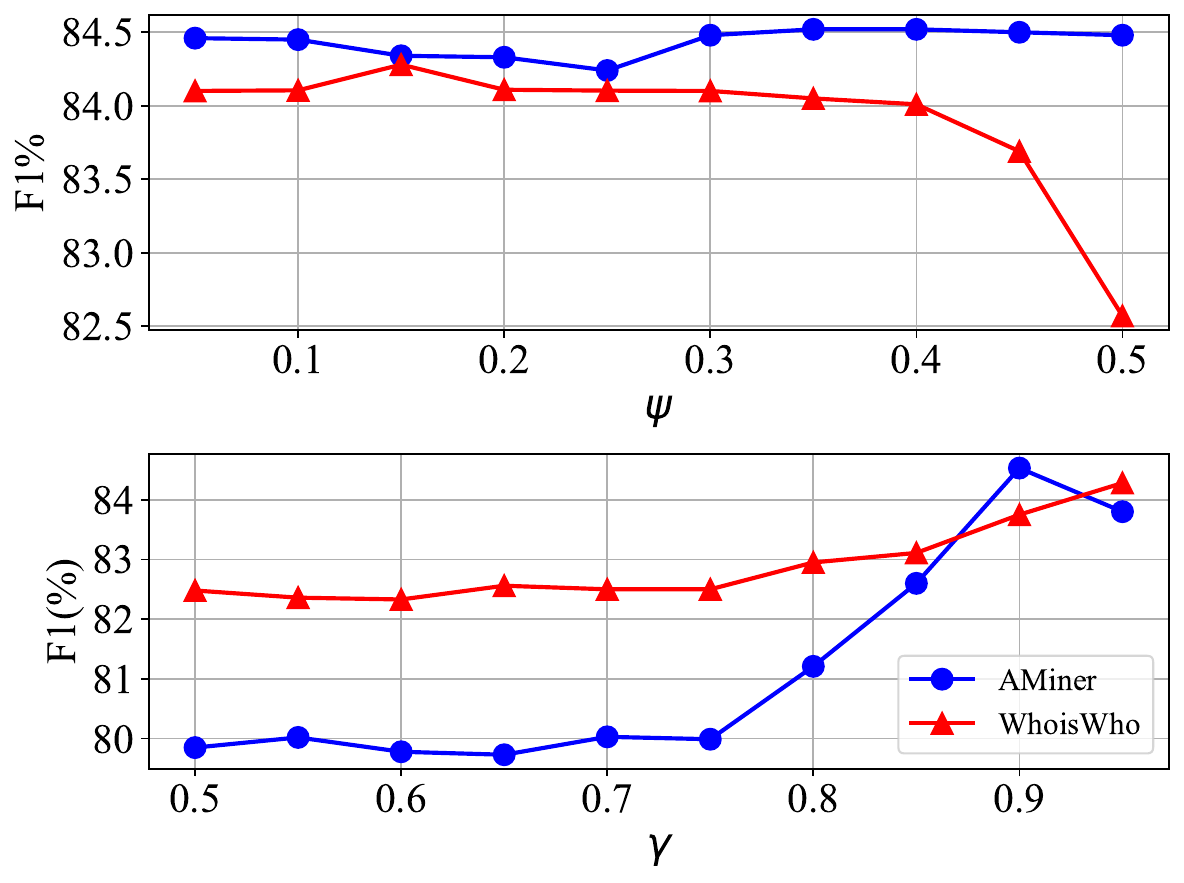}
  \caption{Parameter analysis of $\psi$ and $\gamma$.}
  \label{q1q2}
\end{figure}

Next, we set the range of $\psi$ to 0-0.5 and $\gamma$ to 0.5-0.95, and test the effect of different thresholds in Figure~\ref{q1q2}. It can be seen that $\psi$ does not have a significant effect on the results, while f1 has a significant improvement with the increase of $\gamma$. This result is consistent with ARCC~\cite{liu2024author} using multiple graphical modifications. We believe that text pairs with lower similarity are less likely to have concatenated edges themselves, so $\psi$ has a smaller impact on the results. Embeddings that aggregate more information, on the other hand, are better able to complement correct concatenated edges that should exist, leading to a significant impact of $\gamma$ on performance. In Table~\ref{iteration}, we verified that iterative modification yields lower gains for GRAND-HC. When the number of iterations is too high instead, it leads to a slight decrease in F1, while the complexity of $k$ modifications for $N$ nodes with $d$ dimension feature is $\mathcal{O}(N^{2} \cdot k\cdot d)$. Combining efficiency and performance, we choose to perform graph optimization only once.

\begin{table}
\small\sf\centering
\caption{Analysis of graph refine iterations on F1 score (\%).}
\label{iteration}
 \tabcolsep 0.03in
\begin{tabular}{cccccccc}
\toprule
 Iterations& 0 & 1& 5 & 10 & 15 & 30&60 \\
\midrule
AMiner-v2 &82.83 & 84.86& \textbf{84.89}&84.70&84.82 & 84.73&84.65 \\
WhoisWho-v1 & 81.98 & 84.33& \textbf{84.46}&84.25&84.29 & 84.34&83.24 \\
\bottomrule
\end{tabular}
\end{table}

Finally to show in more detail how the graph-refined distance matrix(GRDM), we visualized it in Figure~\ref{hot}. In Figure~\ref{hot}(a) the ground truth based text pair relations are given as the labeling matrix. The primitive cosine distance matrix is shown in Figure~\ref{hot}(b), which can be seen to be much different from the labeling matrix in Figure~\ref{hot}(a). GRDM firstly optimizes the structure of the graph as Figure~\ref{hot}(d), and adaptively adjusts to Figure~\ref{hot}(c) based on the optimized topology and degree of nodes. Compared with Figure~\ref{hot}(b), its paper pair relationship is more accurate, which enables the HAC clustering algorithm to aggregate effectively, and the F1 score of \textit{`hongtao\_liu'} is improved by nearly 20\%.

\section{Conclusion}
\label{conclusion}
This paper proposes GRAND-HC, a novel SND framework addressing the imbalanced author class distribution via harmony contrastive learning and a graph-refined distance matrix. By dynamically reweighting misclassified pairs, harmony contrastive learning mitigates overemphasis on features of prolific authors while enhancing distinctions for less-published authors. The graph-refined distance matrix further refines clustering by incorporating node connectivity to prevent over-merging. Experiments demonstrate GRAND-HC achieves SOTA performance in macro-F1 among recent SND methods. Besides, GRAND-HC has been deployed in a billion-scale academic system, which validates its scalability and practical efficacy in real-world author disambiguation tasks. This work advances robust AND solutions for scholarly data management. In our upcoming project, we aim to leverage large language models to enhance the accuracy and efficiency of clustering algorithms.

\begin{acks}
This work is funded by NSFC (No. 62525209, T2421002, 623B2071), and Shanghai Pilot Program for Basic Research - Shanghai Jiao Tong University.
\end{acks}

\bibliographystyle{SageH}
\bibliography{sample-base}

\end{document}